\date{}
\documentclass{sig-alternate}

\usepackage{amsmath, graphicx,epsfig ,enumerate,latexsym,amssymb}

\newtheorem{theorem}{Theorem}

\newtheorem{lemma}{Lemma}
\newtheorem{corollary}{Corollary}

\newcommand{\ignore}[1]{}
\newcommand{\Exp}        {\ensuremath{{\bf E}}}

\begin{document}

\conferenceinfo{STOC'12,} {May 19--22, 2012, New York, New York, USA.}
\CopyrightYear{2012}
\crdata{978-1-4503-1245-5/12/05}
\clubpenalty=10000
\widowpenalty = 10000

\iffalse
\title{Competitive Contagion in Networks}
\author{
Sanjeev Goyal\\
Faculty of Economics and Christ's College\\
University of Cambridge
\and
Michael Kearns
\thanks{
Portions of this research
were conducted while the author was visiting the Faculty of Economics and Christ's College, University of
Cambridge.}\\
Computer and Information Science\\
University of Pennsylvania
}
\fi

\title{Competitive Contagion in Networks}
\numberofauthors{2}
\author{
\alignauthor Sanjeev Goyal\\
\affaddr{Faculty of Economics and Christ's College} \\
\affaddr{University of Cambridge} \\
\affaddr{Cambridge, England} \\
\email{sg472@cam.ac.uk}
\alignauthor Michael Kearns
\thanks{
Portions of this research
were conducted while the author was visiting the Faculty of Economics and Christ's College, University of
Cambridge.}\\
\affaddr{Computer and Information Science} \\
\affaddr{University of Pennsylvania} \\
\affaddr{Philadelphia, PA} \\
\email{mkearns@cis.upenn.edu}
}

% \date{\today}
\date{October 28, 2011}

\maketitle
\begin{abstract}

We develop a game-theoretic framework for the study of
competition between firms who have budgets to ``seed'' the initial
adoption of their products by consumers located in a social network.
The payoffs to the firms are the eventual number of adoptions of their product
through a competitive stochastic diffusion process in the network.
This framework
yields a rich class of competitive strategies, which depend in subtle ways
on the stochastic dynamics of adoption, the relative budgets of the players, and the underlying structure of the
social network.

We identify a general property of the adoption dynamics ---  namely, decreasing returns
to local adoption --- for which the inefficiency of resource use at equilibrium (the {\em Price of Anarchy\/})
is uniformly bounded above, across all networks.
We also show that if this property is violated the Price of Anarchy can be unbounded, thus yielding
sharp threshold behavior for a broad class of dynamics.

We also introduce a new notion, the {\em Budget Multiplier\/}, that measures the extent that
imbalances in player budgets can be amplified at equilibrium. We again
identify a general property of the adoption dynamics --- namely, proportional local adoption
between competitors --- for which the (pure strategy) Budget Multiplier is uniformly bounded above,
across all networks. We show that a violation of this property
can lead to unbounded Budget Multiplier, again yielding sharp threshold behavior for a broad
class of dynamics.
\end{abstract}

%A category including the fourth, optional field follows...
% \category{D.2.8}{Software Engineering}{Metrics}[complexity measures, performance measures]

% \terms{Theory}


\section{Introduction}

\noindent
The role of social networks in
shaping individual choices has been brought out in a number of
studies over the years.\footnote{See e.g., Coleman~\cite{col} on
doctors' prescription of new drugs, Conley and Udry~\cite{cu} and
Foster and Rosenzweig~\cite{fr} on farmers' decisions on crop and input
choice, and Feick and Price~\cite{fp}, Reingen et al.~\cite{rfba}, and
Godes and Mayzlin~\cite{gm2} on brand choice by consumers.} In the past,
the deliberate use of such social influences by external agents was
hampered by the lack of good data on social networks. In recent years, data from
on-line social networking sites along with other
advances in information technology have created interest in ways that firms and governments can use
social networks to further their goals.\footnote{The popularity of terms such as
\textit{word of mouth marketing}, \textit{viral marketing},
\textit{seeding the network} and \textit{peer-leading intervention}
is an indicator of this interest.}

In this work, we study competition between firms
who use their resources to maximize product adoption
by consumers located in a social network.
\footnote{Our model may apply to other settings of competitive contagion, such
as between two fatal viruses in a population.}
The social network may transmit information about products, and
adoption of products by neighbors may have direct consumption benefits. The firms,
denoted {\em Red\/} and {\em Blue\/}, know the graph which defines the social network and
offer similar or interchangeable products or services.
The two firms simultaneously choose to allocate their resources
on subsets of consumers, i.e., to \textit{seed} the network with initial adoptions.
The stochastic dynamics of local adoption determine how the influence of each player's
seeds spreads throughout the graph to create new adoptions.
Our work thus builds upon recent interest in models of competitive
contagion~\cite{Kempe,Borodin,Chasparis}.

A distinctive feature of our framework is that we allow for a broad class of local influence processes.
We decompose the dynamics into two parts: a {\em switching function\/} $f$,
which specifies the probability of a consumer switching from non-adoption to adoption as a function of the
fraction of his neighbors who have adopted {\em either\/} of the two products Red and Blue;
and a {\em selection function\/} $g$, which specifies, conditional on switching,
the probability that the consumer adopts (say) Red as function of the fraction of adopting neighbors who have
adopted Red.
Each firm seeks to maximize the total number of consumers who adopt its product.
Broadly speaking, the switching function captures ``stickiness'' of the (interchangeable) products based on their
local prevalence, and the selection function captures preference for firms based on local market share.

This framework yields a rich class of competitive strategies, which depend in subtle ways
on the dynamics, the relative budgets of the players, and the structure of the
social network (Section \ref{sectionexamples} gives some warm-up examples illustrating this point).
Here we focus on understanding
two general features of equilibrium: first, the efficiency of resource use by the players (Price of Anarchy) and
second, the role of the network and dynamics in amplifying ex-ante resource differences between the players (Budget Multiplier).

Our \textit{first} set of results concern efficiency of resource use by the players.
For a fixed graph and fixed local dynamics (given by $f$ and $g$), and budgets of $K_R$ and $K_B$ seed infections for the players,
let $(S_R,S_B)$ be the sets of seed infections that maximize the joint expected infections (payoffs)
$\Pi_R(S_R,S_B) + \Pi_B(S_R,S_B)$ subject to
$|S_R| = K_R$
and
$|S_B| = K_B$, and let
$\sigma_R$ and $\sigma_B$ be Nash equilibrium strategies
obeying the budget constraints that
{\em minimize \/} the joint payoff
$\Pi_R(\sigma_R,\sigma_B) + \Pi_B(\sigma_R,\sigma_B)$
across all Nash equilibria.
The {\em Price of Anarchy\/} (or PoA)\footnote{The PoA is a measure of the maximum potential inefficiency created by non-cooperative/decentralized activity.
In our context, if we suppose that consumers get positive utility from consumption of firms' products then the PoA
also reflects losses in consumer welfare.}
 is then defined as:
\begin{equation*}
\frac{\Pi_R(S_R,S_B) + \Pi_B(S_R,S_B)}
	{\Pi_R(\sigma_R,\sigma_B) + \Pi_B(\sigma_R,\sigma_B)}.
\end{equation*}
Our first main result, Theorem \ref{theoremPOA1}, shows that if the switching function $f$ is concave and the selection function
$g$ is linear, then the PoA is uniformly bounded above by 4, across all networks.
The main proof technique we employ is the construction of certain coupled stochastic dynamical processes that
allow us to demonstrate that, under the assumptions on $f$ and $g$, the departure of one player can only
benefit the other player, even though the total number of joint infections
can only decrease.
This in turn lets us argue that players can successfully defect to the maximum social
welfare solution and realize a significant fraction of its payoff, thus implying they must also do so at equilibrium.

Our next main result, Theorem \ref{theoremPOA2}, shows that even a small amount of convexity in the
switching function $f$ can lead to arbitrarily high PoA. This result is obtained by
constructing a family of layered networks whose dynamics effectively compose $f$ many times, thus amplifying its convexity.
Equilibrium and large PoA are enforced by the fact that despite this amplification, the players are better off
playing near each other: this means that if one player locates in one part of the network, the other player
has an incentive to locate close by, even if they would jointly be better off locating in a different part of the network.
Taken together, our PoA upper and lower bounds yield sharp threshold behavior in a parametric classes of dynamics.
 For example, if the switching function is $f(x) = x^r$ for $r > 0$ and the
selection function $g$ is linear, then for all $r \leq 1$ the PoA is at most 4, while for any $r > 1$ it can
be unbounded.\footnote{\label{hhnote} The Price of Stability (PoS) compares outcomes in the `best' equilibrium with socially optimal
outcomes; one may interpret the PoS as a measure of the minimum inefficiency created by non-cooperative, as opposed
to merely decentralized, activity.
Heidari~\cite{hh}, adapting the proofs presented here, has recently shown that as with PoA,
even slight convexity of the switching function is sufficient to
generate arbitrarily large PoS.}

Our \textit{second} set of results are about the effects of networks and dynamics on
budget differences across the players. We introduce and study a new quantity called the {\em Budget Multiplier\/}.
For any fixed graph, local dynamics, and initial budgets, with $K_R \geq K_B$, let $(\sigma_R,\sigma_B)$
be the Nash equilibrium that {\em maximizes\/} the quantity
\begin{equation*}
\frac{\Pi_R(\sigma_R, \sigma_B)}
    {\Pi_B(\sigma_R, \sigma_B)}
\times \frac{K_B}{K_R}
\end{equation*}
among all Nash equilibria $(\sigma_R,\sigma_B)$; this quantity is just the ratio of
the final payoffs divided by the ratio of the initial budgets. The resulting maximized quantity
is the Budget Multiplier, and it measures the extent
to which the larger budget player can obtain a final market share
that exceeds their share of the initial budgets.

Theorem \ref{theoremPOB1} shows that if the switching function is concave and the
selection function is linear, then the (pure strategy) Budget Multiplier is bounded above by 2, uniformly across
all networks. The proof imports elements of the proof for the PoA upper bound, and additionally
employs a method for attributing adoptions back to the initial seeds that generated them.

Our next result, Theorem \ref{theoremPOB2}, shows that even a slight departure from linearity
in the selection function can yield unbounded Budget Multiplier.
The proof again appeals to network structures that amplify the nonlinearity of $g$ by self-composition,
which has the effect of ``squeezing out'' the player with smaller budget.
Combining the Budget Multiplier upper and lower bounds again allows us to exhibit simple parametric forms yielding
threshold behavior: for instance, if $f$ is linear and $g$ is from the well-studied Tullock contest function
family (discussed shortly), which includes linear $g$ and therefore bounded Budget Multiplier, even an infinitesimal departure
from linearity can result in unbounded Budget Multiplier.

\noindent\textit{\bf Related Literature.}
Our paper contributes to the study of competitive strategy in network environments.
We build a framework which combines ideas from economics (contests, competitive seeding and advertising)
and computer science -- uniform bounds on properties of equilibria, as in the Price of Anarchy --
to address a topical and natural question. The Tullock contest function was introduced in Tullock~\cite{tullock2};
for an axiomatic development see Skaperdas~\cite{skaperdas}. For early and
influential studies of competitive advertising, see Butters~\cite{butters} and Grossman and Shapiro~\cite{gs} .
The Price of Anarchy (PoA) was introduced by
Koutsoupias and Papadimitriou~\cite{katpapa}, and important early results bounding the PoA in networked settings
regardless of network structure were given by
Roughgarden and Tardos~\cite{roughgarden}.
The tension between equilibrium and Nash efficiency is a recurring theme in
economics; for a general result on the inefficiency of Nash equilibria, see Dubey~\cite{dubey}.

More specifically, we contribute to the study of influence in networks.
This has been an active field of study in the last decade, see e.g.,
Ballester, Calvo-Armengol and Zenou~\cite{ref:ncz}; Bharathi, Kempe and Salek~\cite{Kempe};
Galeotti and Goyal~\cite{galgoyal}; Kempe, Kleinberg, and Tardos~\cite{kkt1,kkt2};
Mossel and Roch~\cite{Mossel};
Borodin, Filmus, and Oren~\cite{Borodin};
Chasparis and Shamma~\cite{Chasparis}; Carnes et al~\cite{carnes};
Dubey, Garg and De Meyer~\cite{dubey2}; Vetta~\cite{vetta}.
There are three elements in our framework which appear to be novel:
one, we consider a fairly general class of adoption rules at
the individual consumer level which correspond to different roles
which social interaction can potentially play (existing work often considers specific local dynamics);
two, we study competition for influence in a network (existing work has often focused on the case of
a single player seeking to maximize influence), and three, we introduce and study the notion of Budget Multiplier as a measure of how
networks amplify budget differences. To the best of our knowledge, our results on the relationship between
the dynamics and qualitative features of the strategic equilibrium are novel. Nevertheless, there are definite
points of contact between our results and proof techniques and earlier research in (single-player and competitive)
contagion in networks that we shall elaborate on where appropriate.

\section{Model}

\subsection{Graph, Allocations, and Seeds}

We consider a 2-player game of competitive adoption on a
(possibly directed) graph $G$ over $n$ vertices. $G$ is known
to the two players, whom we shall refer to as $R$(ed) and $B$(lue).\footnote{The restriction to
2 players is primarily for simplicity; our main result on PoA can be generalized to a game with 2 or more players,
see statement in Section~\ref{sec:PoAUB} below.}
We shall also use $R, B$ and $U$(ninfected) to denote the state of
a vertex in $G$, according to whether it is currently infected by one of the
two players or uninfected. The two players
simultaneously choose some number of vertices to initially seed;
after this seeding, the stochastic dynamics of local adoption (discussed below)
determine how each player's seeds spread throughout $G$ to
create adoptions by new nodes. Each player seeks to maximize
their (expected) total number of eventual adoptions.
\footnote{Throughout the paper, we shall use the terms {\em infection\/} and {\em adoption\/} interchangeably.}

More precisely, suppose that player $p=R,B$ has {\em budget\/} $K_p\in \mathbf{N}_+$;
Each player $p$ chooses an allocation of budget across the $n$ vertices, $a_p=(a_{p1},
a_{p2},...,a_{pn})$, where  $a_{pj}\in\mathbf{N}_+$ and
$\sum_{j=1}^{n} a_{pj} = K_p$. Let $A_p$ be the set of allocations for
player $p$, which is their pure strategy space.
A mixed strategy for player $p$ is a probability
distribution $\mathcal{\sigma}_p$ on $A_p$. Let $\mathcal{A}_p$
denote the set of probability distributions for player $p$.
The two players simultaneously choose their strategies $(\sigma_R,\sigma_B)$. Consider any
realized initial allocation $(a_R,a_B)$ for the two players. Let
$V(a_R)=\{v|a_{vR}>0\}$, $V(a_B)=\{v|a_{vB}>0\}$ and let
$V(a_R,a_B)=V(a_R)\cup V(a_B)$. A vertex $v$ becomes initially infected if one or
more players assigns a seed to infect $v$.
If both players assign seeds to the same vertex, then the probability of initial
infection by a player is proportional to the seeds allocated by
the player (relative to the other player). More precisely,
fix any allocation $(a_R, a_B)$.
For any vertex $v$, the initial state $s_v$ of $v$ is in $\{R,B\}$
if and only if $v\in V(a_R,a_B)$.  Moreover,
$s_{v}= R$ with  probability $a_{vR}/(a_{vR}+a_{vB})$, and
$s_{v}= B$ with probability $a_{vB}/(a_{vR}+a_{vB})$.

Following the allocation of seeds, the stochastic contagion
process on $G$ determines how these $R$ and $B$ infections generate new adoptions in the network.
We consider a discrete time model for this process.
The state of a vertex $v$ at time $t$ is denoted $s_{vt}\in\{U, R, B\}$, where $U$ stands for Uninfected, $R$
stands for infection by $R$, and $B$ stands for infection by $B$.
% Occasionally, we will refer to a state in $\{R,B\}$ as
% $I$(nfected) without specifying by which player.
% The state of the network $G$ at time $t$
% is $s_t$.

\subsection{The Switching-Selection Model}
\label{sec:contsec}

We assume there is an {\em update schedule\/} which determines the
order in which vertices are considered for state updates. The primary
simplifying assumption we shall make about this schedule is that once
a vertex is infected, it is never a candidate for updating again.
% \mk{Why and where do we really need/use this? SG this seems like a natural assumption in a model in which firms seek to
%maximize influence. If nodes may get uninfected or switch colors then payoffs we have written down are not well defined.}
%\footnote{This assumption can be relaxed considerably, at the expense of complicating some of our proofs.}

Within this constraint, we allow for a variety of behaviors, such as
randomly choosing an uninfected vertex to update at each time step (a form of
{\em sequential\/} updating), or updating all uninfected
vertices simultaneously at each time step (a form of {\em parallel\/} updating).
We can also allow for an
{\em immunity\/} property --- if a vertex is exposed once to infection and remains
uninfected after updating, it is never updated again.
Update schedules may also have finite termination
times or conditions --- for instance,
if the firms primarily care about the number of adoptions
in the coming fiscal year.
We can also allow schedules that update each uninfected vertex only a fixed number of times.
Note that a schedule
which perpetually updates uninfected vertices will eventually cause any connected $G$
to become entirely infected, thus trivializing the PoA (though not necessarily the
Budget Multiplier), but we allow for considerably more general schedules.
\footnote{The proof of Lemma~\ref{lem:coupled} specifies the technical property we need of
the update schedule, which is consistent with the examples mentioned here and many others.}

For the stochastic update of an uninfected vertex $v$, we will
consider what we shall call the {\em switching-selection\/} model.
In this model, updating is determined by the application
of two functions to $v$'s local neighborhood:
$f(x)$ (the {\em switching\/} function), and
$g(y)$ (the {\em selection\/} function).
More precisely, let $\alpha_R$ and $\alpha_B$ be the fraction of $v$'s
neighbors infected by $R$ and $B$, respectively, at the time of the
update, and let $\alpha = \alpha_R + \alpha_B$ be the total fraction
of infected neighbors. The function $f$ maps $\alpha$ to the interval
$[0,1]$ and $g$ maps $\alpha_R/(\alpha_R + \alpha_B)$ (the relative
fraction of infections that are $R$) to $[0,1]$. These two functions
determine the stochastic update in the following fashion:
\begin{enumerate}
\item With probability $f(\alpha)$, $v$ becomes infected by {\em either\/} $R$ or $B$;
    with probability $1 - f(\alpha)$, $v$ remains in state $U$(ninfected), and the update ends.
\item If it is determined that $v$ becomes infected, it becomes infected by $R$
    with probability
\[
g(\alpha_R/(\alpha_R+\alpha_B)),
\]
and infected by $B$
with probability
\[
g(\alpha_B/(\alpha_R+\alpha_B)).
\]
\end{enumerate}
We assume $f(0) = 0$ (infection requires exposure),
$f(1) = 1$ (full neighborhood infection forces infection),
and $f$ is increasing (more exposure yields more infection);
and $g(0) = 0$ (players need some local market share to win an infection),
$g(1) = 1$. Note that since the selection step above requires that an infection
take place, we also have $g(y) + g(1-y) = 1$, which implies $g(1/2) = 1/2$.
We assume that the switching and selection functions are the same across vertices.
\footnote{This is for expositional simplicity only; our main results on PoA and Budget Multiplier carry
over to a setting with heterogeneity across vertices (so long as the selection function remains symmetric across
the two players).}

We think of the switching function as specifying how rapidly adoption increases
with the fraction of neighbors who have adopted (i.e. the stickiness of the interchangeable
products or services), regardless of their $R$ or $B$ value; while the selection function specifies the probability of infection by each firm
in terms of the local relative market share split.\footnote{In the threshold model a consumer switches to an action once a certain fraction
of society/neighborhood adopts that action (Granovetter, 1978). In our model, heterogeneous thresholds can be captured in terms
of different switching function $f$.} In addition to being a natural decomposition of the dynamics, our results will show that we can articulate
properties of $f$ and $g$ which sharply characterize the PoA and Budget Multiplier.
In Section \ref{sectionexamples}, we shall provide economic motivation for
this formulation and also illustrate with specific parametric families of functions
$f$ and $g$. We also discuss more general models for the local dynamics at a number
of places in the paper. The Appendix also illustrates how these switching and selection functions $f$-$g$ may arise out of
optimal decisions made by consumers located in social networks.

\noindent
\textit{\bf Relationship to Other Models.\/} It is natural to consider
both general and specific relationships between our models and others in the literature,
especially the widely studied {\em general threshold\/}
model~\cite{kkt1,Mossel,Kempe}. One primary difference is our allowance of rather general
choices for the switching and selection functions $f$ and $g$, and our study of how these
choices influence equilibrium properties. When considering concave $f$ --- which is
a special case of sub-modularity --- the relationship becomes closer, and our proof techniques
bear similarity to those in the general threshold model (particularly the extensive use of
coupling arguments). Nevertheless there seem to be elements of our model not easily captured
in the general threshold model, including our allowance of rather general update schedules
that may depend on the state of a vertex; the general threshold model asks that all randomization
(in the form of the selection of a random threshold for each vertex) occur prior to the updating process,
whereas our model permits repeated randomization in subsequent updates, in a possibly state-dependent
fashion. We shall make related technical comments where appropriate.

\subsection{Payoffs and Equilibrium}

Given a graph $G$ and an initial allocation of seeds $(a_R, a_B)$, the dynamics described above --- determined
by $f$, $g$, and the update schedule --- yield a stochastic number of
eventual infections for the two players.
For $p=R,B$, let $\chi_p$
denote this random variable for $R$ and $B$,
respectively, at the termination of the dynamics. Given strategy
profile $(\sigma_R,\sigma_B)$, the payoff to player $p=R, B$ is
$\Pi_p(\sigma_R, \sigma_B)=\Exp[\chi_p|(\sigma_R,\sigma_B)]$.
Here the expectation is over any randomization in the player strategies in the choice of
initial allocations, and the randomization in the stochastic updating dynamics.
A Nash equilibrium is a profile of strategies $(\sigma_R,\sigma_B)$ such that
$\sigma_p$ maximizes player $p$'s payoff given the strategy
$\sigma_{-p}$ of the other player.

\subsection{Price of Anarchy and Budget Multiplier}

For a fixed graph $G$, stochastic update dynamics, and
budgets $K_R, K_B$, the {\em maximum payoff\/} allocation is the (deterministic) allocation
$(a_R^*,a_B^*)$ obeying the budget constraints that maximizes
$\Exp[\chi_R + \chi_B|(a_R,a_B)]$.
For the same fixed graph, update dynamics and budgets, let
$(\sigma_R,\sigma_B)$ be the Nash equilibrium strategies that {\em minimize\/}
$\Exp[\chi_R + \chi_B|(\sigma_R,\sigma_B)]$ among all Nash equilibria
--- that is, the Nash equilibrium with the smallest joint payoff. Then
the {\em Price of Anarchy\/} (or PoA) is defined to be
\begin{equation*}
\frac{\Exp[\chi_R + \chi_B|(a_R^*,a_B^*)]}
    {\Exp[\chi_R + \chi_B|(\sigma_R,\sigma_B)]}
\end{equation*}
The Price of Anarchy is a measure of the inefficiency in resource use created due to decentralized/
non-cooperative behavior by the two players.
In the context of competition between firms, one interpretation of the PoA
is as a measure of the relative improvement in efficiency effected by a hypothetical
merger of the firms.

We also introduce and study a new quantity called the {\em
Budget Multiplier\/}. The Budget Multiplier measures the extent to which network structure and
dynamics can amplify initial resource inequality across the players. Thus
for any fixed graph $G$ and stochastic update dynamics, and initial
budgets $K_R, K_{B}$, with $K_R\geq K_{B}$, let $(\sigma_R,\sigma_B)$ be the Nash
equilibrium that {\em maximizes\/} the ratio
\begin{equation*}
\frac{\Pi_R(\sigma_R, \sigma_B)}
    {\Pi_B(\sigma_R, \sigma_B)}
\times \frac{K_B}{K_R}
\end{equation*}
among all Nash equilibria.
The resulting maximized ratio is the Budget Multiplier, and it measures the extent
to which the larger budget player can obtain a final market share
that exceeds their share of the initial
budgets.

\section{Local Dynamics: Motivation}\label{sectiondynamics}

In this section, we provide some
examples of the
decomposition of the local update dynamics into a switching function $f$
and a selection function $g$. As discussed above, we view the switching function
as representing how contagious a product or service is, regardless of
which competing party provides it; and we view the selection function as
representing the extent to which a firm having majority local market share
favors its selection in the case of adoption.
We illustrate the richness of this model by
examining a variety of different mathematical choices for the functions $f$ and $g$, and
discuss examples from the domain of technology adoption that might (qualitatively)
match these forms. Finally, to illustrate the scope of this formulation,
we also discuss examples of natural update dynamics that {\em cannot\/} be decomposed in this way.

\begin{figure*}[tbh]
 \begin{center}
  \includegraphics[scale=.4]{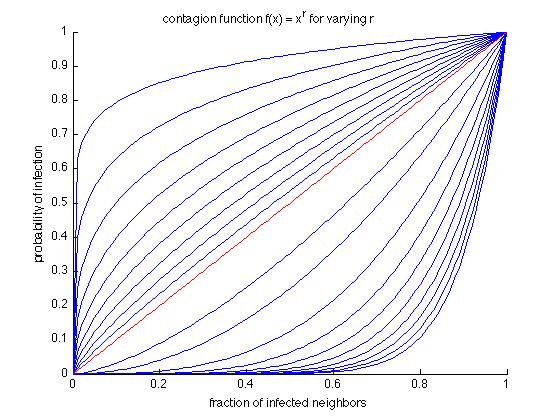}
 \includegraphics[scale=.4]{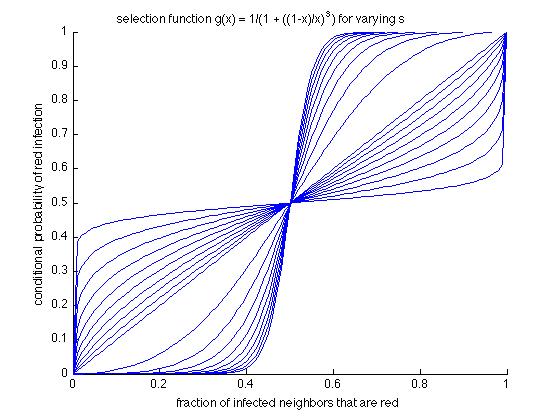}
\end{center}
\caption{Left: Plots of $f(x) = x^r$ for
     varying choices of $r$, including $r=1$ (linear, red line),
    $r < 1$ (concave), and $r > 1$ (convex).
   Right: Plots of $g(y) = y^s/(y^s+(1-y)^s)$ for varying choices of $s$,
  including $s=1$ (linear, red line),
  $s < 1$ (equalizing), and $s > 1$ (polarizing).
}
     \label{fig:decomposition}
\end{figure*}

A fairly broad class of dynamics is captured by the following parametric family of functions.
The switching function
\begin{equation*}
f(x) = x^r\;\; r \geq 0
\end{equation*}
and the selection function
\begin{equation*}
g(y) = y^s/(y^s + ((1-y)^s)
\;\; s \geq 0.
\end{equation*}

Regarding this form for $f$, for $r=1$ we have linear adoption.
For $r < 1$ we have $f$ concave, corresponding to cases in which the
probability of adoption rises quickly with only a small fraction of adopting
neighbors, but then saturates or levels off with larger fractions of adopting
neighbors. In contrast, for $r > 1$ we have $f$ convex, which at very large
values of $r$ can begin to approximate threshold adoption behavior --- the
probability of adoption remains small until some fraction of neighbors
has adopted, then rises rapidly. See Figure~\ref{fig:decomposition}.

Regarding this form for $g$, which is known as the {\em Tullock contest function\/} (Tullock (1980)),
for $s=1$ we have a (linear) {\em voter\/} model in which the probability of selection is proportional to local
market share. For $s < 1$ we have what we shall call an {\em equalizing\/} $g$, by which we
mean that selection of the minority party in the neighborhood is favored
relative to the linear voter model $g(y) = y$; and for $s > 1$ we have
a {\em polarizing\/} $g$, meaning that the minority party is disfavored
relative to the linear model. As $s$ approaches 0, we approach the completely
equalizing choice $g \equiv 1/2$, and as $s$ approaches infinity, we
approach the completely polarizing {\em winner-take-all\/} $g$;
see Figure~\ref{fig:decomposition}.

These parametric families of switching and selection functions will play an important
role in illustrating our general results. We now discuss a few technology adoption examples which
are (qualitatively) covered by these families of functions.
\begin{itemize}
\item {\em Social Network Services (Facebook, Google+, etc.):\/}
Here adoption probabilities might grow slowly with a small fraction
of adopting neighbors, since there is little value in using (any) social networking
services if none of your friends are using them; thus a convex switching function $f$ ($r > 1$) might be a
reasonable model. However, given that it is currently difficult or impossible
to export friends and other settings from one service to another, there are
strong platform effects in service selection, so a polarizing or even winner-take-all
selection function $g$ ($s > 1$) might be most appropriate.
\item {\em Televisions (Sharp, Sony, etc.):\/}
Televisions were immediately useful upon their
introduction,
without the need for adoption by neighbors, since they allowed immediate access to broadcast programming;
the adoption by neighbors serves mainly as a route for information sharing
about value of the product. The information value of more neighbors adopting a product is falling with adoption
and so a concave $f$ might be appropriate.
Compared to social networking services, the platform effects are lower here, and so a linear
or equalizing $g$ is appropriate.
\item {\em Mobile Phone Service (Verizon, T-Mobile, etc.):\/}
Mobile phone service was immediately useful upon its introduction without adoption
by neighbors, since one could always call land lines, thus arguing for a concave
$f$. Since telephony systems need to be interoperable,
platform effects derive mainly from marketing efforts such as ``Friends and Family''
programs, and thus are extant but perhaps weak, suggesting an equalizing $g$.
\end{itemize}

In the proofs of some of our results, it will sometimes be
convenient to use a more general adoption function formulation with
some additional technical conditions that are met by our switching-selection
formulation. We will refer to this general, single-step model
as the {\em generalized adoption function\/} model. In this model,
if the local fractions of Red and Blue neighbors are $\alpha_R$ and
$\alpha_B$, the probability that we update the vertex with an $R$
infection is $h(\alpha_R,\alpha_B)$ for some {\em adoption
function\/} $h$ with range $[0,1]$, and symmetrically the
probability of $B$ infection is thus $h(\alpha_B,\alpha_R)$. Let us
use $H(\alpha_R,\alpha_B) = h(\alpha_R,\alpha_B) +
h(\alpha_B,\alpha_R)$ to denote the total infection probability
under $h$. Note that we can still always decompose $h$ into a
two-step process by defining the switching function to be
$f(\alpha_R,\alpha_B) = H(\alpha_R,\alpha_B)$ and defining the
selection function to be
\[
g(\alpha_R,\alpha_B) =
h(\alpha_R,\alpha_B)/(h(\alpha_R,\alpha_B) + h(\alpha_B,\alpha_R))
\]
which is the infection-conditional probability that $R$ wins the infection.
The switching-selection model is thus the special case of the
generalized adoption function model in which $H(\alpha_R,\alpha_B) = f(\alpha_R+\alpha_B)$
is a function of only $\alpha_R + \alpha_B$, and
$g(\alpha_R,\alpha_B)$ is a function of only $\alpha_R/(\alpha_R +
\alpha_B)$.\footnote{While the decomposition in terms of a switching function and a selection function
accommodates a fairly wide range of adoption dynamics there are some
cases which are ruled out. Consider the choice $h(x,y) = x(1-y^2)$; it is
easily verified that
the total probability of adoption $H(x,y)$ is increasing in $x$ and $y$.
But $H(x,y)$ clearly cannot be expressed as a function of the form $f(x+y)$. Similarly, it is easy to construct an
adoption function that is not only not decomposable, but violates monotonicity.
Imagine consumers that prefer to adopt the majority choice in their neighborhood, but
will only adopt once their local neighborhood market is sufficiently
settled in favor of one or the other product.
The probability of total adoption may then be higher with $x=0.2$, $y=0$
as compared to $x=y=0.4$.}

\section{Equilibrium Examples}\label{sectionexamples}

The examples here illustrate that our framework yields a rich class of competitive strategies, which can depend in subtle ways
on the dynamics, the relative budgets of the players and the structure of the
social network.

\textbf{Price of Anarchy:} Suppose that budgets of the
firms are $K_R=K_B=1$, and the update rule is such that all vertices
are updated only once. The network contains two
connected components with $10$ vertices and $100$ vertices,
respectively. In each component there are 2 influential vertices,
each of which is connected to the other 8 and 98 vertices,
respectively. So in component $1$, there are $16$ directed links
while in component $2$ there are $196$ directed links in all.

\begin{itemize}
\item Suppose that the switching function and the selection function are
both linear, $f(x)=x$ and $g(y)=y$. Then there is a unique
equilibrium in which players place their seeds on
distinct influential vertices of component 2. The total infection is
then $100$ and this is the maximum number of infections possible
with $2$ seeds. So here the PoA is $1$.

\item Let us now alter the switching function such that $f(1/2)=\epsilon$ for some
$\epsilon<1/2$ (keeping $f(1)=1$, as always), but retain the
selection function to be $g(y)=y$. Now there also exists an equilibrium in which
the firms locate on the influential vertices of component $1$. In
this equilibrium payoffs to each player are equal to $5$. Observe that for $\epsilon < 1/25$, a
deviation to the other component is not profitable: it yields an
expected payoff equal to $\epsilon\times 100$, and this is strictly
smaller than $5$.
Since it is still
possible to infect component $2$ with $2$ seeds, the
PoA is $10$. Here inefficiency is created by a coordination
failure of the players.

\item Finally, suppose there is only one component with $110$ vertices,
with $2$ influential vertices and $108$ vertices receiving directed links.
Then equilibrium under both switching functions considered above
will involve firms
locating at the $2$ influential vertices and this will lead to
infection of all vertices. So the PoA is $1$,
irrespective of whether the switching function is linear $f(x)=x$ or
whether $f(1/2)<1/25$.

\end{itemize}

Thus for a fixed network, updating rule and selection
function, variations in the switching function can generate large
variations in the PoA. Similarly, for fixed update rule
and switching and selection functions, a change in the network structure yields
very different PoA.

Theorem \ref{theoremPOA1} provides a set of sufficient conditions on
switching and selection function, under which the PoA
is uniformly bounded from above. Theorem \ref{theoremPOA3} shows how
even small violations of these conditions can lead to arbitrarily
high PoA.
\bigskip

\textbf{Budget Multiplier:} Suppose that budgets of the firms are
$K_R=1$, $K_B=2$ and the update rule is such that all vertices are
updated only once. The network contains $3$ influential
vertices, each of which has a directed link to all the other $n-3$
vertices, respectively. So there are $3(n-3)$ links in all. Let
$n\gg3$.

\begin{itemize}
\item Suppose the switching function and selection function are both
linear, i.e., $f(x)=x$ and $g(y)=y$. There is a unique equilibrium
and in this equilibrium, players will place their resources on
distinct influential vertices. The (expected) payoffs to player $R$
are $n/3$, while the payoffs to player $B$ are $2n/3$. So the
Budget Multiplier is equal to $1$.

\item Next, suppose the switching function is convex with $f(2/3)=1/25$,
and the selection function $g(y)$ is as in Tullock (1980). Suppose the two
players place their resources on the three influential vertices. The
payoffs to $R$ are $g(1/3)n$, while firm $B$ earns $g(2/3)n$.
Clearly this is optimal for firm $B$ as any deviation can only lower
payoffs. And, it can be checked that a deviation by firm $R$ to one
of the influential vertices occupied by player $B$ will yield a payoff
of $n/100$ (approximately). So the configuration specified is an
equilibrium so long as $g(1/3)\geq 1/100$. The Budget Multiplier is
now (approximately) 50.

\item Finally, suppose the network consists of $\ell$ equally-sized connected
components. In each component, there is $1$ influential vertex which
has a directed link to each of the $(n/\ell)-1$ other vertices. In
equilibrium each player locates on a distinct influential vertex,
\textit{irrespective} of whether the switching function is convex or
concave and whether the Tullock selection function is linear
($s=1$) or whether it is polarizing ($s>1$). The Budget Multiplier is now equal to 1.

\end{itemize}

These examples show that for fixed network and updating rule, variations
in the switching and selection functions generate large variations
in Budget Multiplier. Moreover, for fixed switching and selection functions the
payoffs depend crucially on the network.

Theorem \ref{theoremPOB1} provides a set of sufficient conditions on
the switching and selection function, under which the Budget Multiplier is
uniformly bounded. Theorem \ref{theoremPOB2} shows how even small
violations of these conditions can lead to arbitrarily high Budget Multiplier.
Theorem \ref{theoremPOB3} illustrates the role of concavity of the
switching function in shaping the Budget Multiplier.

\section{Results: Price of Anarchy}\label{sectionPOA}

We first state and prove a theorem providing general conditions in
the switching-selection model under which the Price of Anarchy is
bounded by a constant that is independent of the size and structure
of the graph $G$. The simplest characterization is that $f$ being
any concave function (satisfying $f(0) = 0$, $f(1) = 1$ and $f$
increasing), and $g$ being the linear voter function $g(y) =
y$ leads to bounded PoA; but we shall see the conditions allow for
certain combinations of concave $f$ and nonlinear $g$ as well. We
then prove a lower bound showing that the concavity of $f$ is
required for bounded PoA in a very strong sense --- a small amount of convexity
can lead to unbounded PoA.

\subsection{PoA: Upper Bound}
\label{sec:PoAUB}

We find it useful to state and  prove our theorems using the generalized adoption model
formulation described in section \ref{sectiondynamics}, but with some additional conditions on $h$ that we now discuss.
If $h(\alpha_R,\alpha_B)$ (respectively, $h(\alpha_B,\alpha_R)$)
is the probability that a vertex with fractions $\alpha_R$ and $\alpha_B$ of $R$ and $B$
neighbors is infected by $R$ (respectively, $B$), we
say that the total infection probability $H(\alpha_R,\alpha_B) = h(\alpha_R,\alpha_B) + h(\alpha_B,\alpha_R)$ is
{\em additive in its arguments\/} (or simply {\em additive\/}) if $H$ can be written $H(\alpha_R,\alpha_B) = f(\alpha_R+\alpha_B)$ for
some increasing function $f$ --- in other words, $h$ permits
interpretation as a switching function.
We shall say that $h$ is {\em competitive\/} if $h(\alpha_R,\alpha_B) \leq h(\alpha_R,0)$ for all
$\alpha_R, \alpha_B \in [0,1]$. In other words, a player always has equal or higher
infection probability in the absence of the other player.

\noindent
{\bf Concave $f$ and linear $g$.\/}
Observe that the switching-selection formulation always satisfies the
additivity property by definition. Moreover, in the switching-selection formulation, if $g$ is linear,
the competitiveness condition becomes
\[
h(x,y) = f(x+y)(x/(x+y)) \leq f(x) = h(x,0)
\]
 or
\[
f(x+y)/(x+y) \leq f(x)/x.
\]
This condition is satisfied by the
concavity of $f$. We will later see that the following theorem also applies to
certain combinations of concave $f$ and nonlinear $g$.
The first theorem can now be stated.

\begin{theorem}\label{theoremPOA1}
If the adoption function $h(\alpha^R,\alpha^B)$ is competitive
and $H$ is additive in its arguments, then
Price of Anarchy is at most 4 for any graph $G$.
\footnote{This result can be generalized to $p\geq 2$ players: In the $p\geq 2$ player game, if $f$ is concave and
$g$ is linear then the PoA is bounded above by $2p$~\cite{hh}.}
\end{theorem}

\noindent
\proof  We establish the theorem via a series of lemmas and inequalities that can be
summarized as follows. Let $(S_R^*,S_B^*)$ be an initial allocation of infections that gives the maximum joint
payoff, and let $(S_R,S_B)$ be a pure\footnote{The extension to mixed strategies is straightforward and omitted.}
Nash equilibrium with $S_R$ being the
larger set of seeds, so $K_R = |S_R^*| = |S_R| \geq K_B = |S_B^*| = |S_B|$.
We first establish a general lemma (Lemma~\ref{lem:coupled}) that implies
that the set $S_R^*$ alone (without $S_B^*$ present) must yield payoffs close to the
maximum joint payoff (Corollary~\ref{corr:coupled}).
The proof involves the construction of a coupled stochastic process technique
we employ repeatedly in the paper.
\footnote{The theorem includes concave
(and therefore sub-modular) $f$ and makes extensive use of coupling arguments to
prove local-to-global effects (of which Lemmas~\ref{lem:coupled} and~\ref{lem:coupled2} are examples);
this bears a similarity to the work of Mossel and Roch~\cite{Mossel}, and it has been
suggested that our proofs might be simplified by appeal to their results. However,
we have not been able to apply their results in our context. Two features of our framework
seem to make direct application difficult: first, the important role of competitive effects, which is explicit
in Lemma~\ref{lem:coupled}; and second, 
the variety of updating schedules we consider 
appear not be covered by the general threshold model which underlies the Mossel and
Roch analysis. While we suspect more direct relationships might be possible in special
cases, here we provide proofs specific to our model.} 
We then contemplate a deviation by
the Red player to $(S_R^*,S_B)$. Another coupling argument
(Lemma~\ref{lem:coupled2}) establishes that the total payoffs for both players under $(S_R^*,S_B)$
must be at least those for the Red player alone under
$(S_R^*,\emptyset)$.
This means that under
$(S_R^*,S_B)$, one of the two players must be approaching the maximum joint infections.
If it is Red, we are done, since Red's equilibrium payoff must also be this large.
If it is Blue, Lemma~\ref{lem:coupled} implies
that Blue could still get this large payoff even after the departure of Red. Next we invoke
Lemma~\ref{lem:coupled2} to show that total eventual payoff to both players under $(S_R,S_B)$ must exceed this large payoff
accruing to Blue, proving the theorem.

\begin{lemma} \label{lem:coupled}
Let $A_R$ and $A_B$ be any sets of seed vertices for the two players. Then if $h$
is competitive and $H$ is additive,
\[
\Exp[\chi_{R}|(A_R,\emptyset)]
    \geq \Exp[\chi_{R}|(A_R,A_B)]
\]
and
\[
\Exp[\chi_{B}|(\emptyset,A_B)]
    \geq \Exp[\chi_{B}|(A_R,A_B)].
\]
\end{lemma}

\noindent
\proof
We provide the proof for the first statement involving $\chi_R$; the proof for $\chi_B$ is
identical.
We introduce a simple {\em coupled simulation\/} technique that we shall appeal to several
times throughout the paper. Consider the stochastic dynamical process on $G$ under two different
initial conditions: both $A_R$ and $A_B$ are present (the {\em joint\/} process, denoted
$(A_R,A_B)$ in the conditioning in the statement of the lemma); and only the set $A_R$ is present
(the {\em solo Red\/} process, denoted $(A_R,\emptyset)$).
Our goal is to define a new
stochastic process on $G$, called the {\em coupled process\/}, in
which the state of each vertex $v$ will be a pair $<X_v,Y_v>$.
We shall arrange that $X_v$ faithfully represents the state of a vertex in the joint process,
and $Y_v$ the state in the solo Red process.
However, these state components will be correlated or a coupled in a deliberate manner.
More precisely, we wish to arrange the coupled process to have the following properties:
\begin{enumerate}
\item \label{prop1} At each step, and for any vertex state $<X_v,Y_v>$,
    $X_v \in \{U,R,B\}$ and
    $Y_v \in \{U,R\}$.
\item \label{prop2} Projecting the states of the coupled process onto either component
    faithfully yields the respective process. Thus, if
    $<X_v,Y_v>$ represents the state of vertex $v$ in the coupled process, then
    the $\{X_v\}$ are stochastically identical to the joint process,
    and the $\{Y_v\}$ are stochastically identical to the solo Red process.
\item \label{prop3} At each step, and for any vertex state $<X_v,Y_v>$,
    $X_v = R$  implies
    $Y_v = R$.
\end{enumerate}
Note that the first two properties are easily achieved by simply running {\em independent\/}
joint and solo Red processes. But this will violate the third property, which
yields the lemma, and thus we introduce the coupling.

For any vertex $v$, we define its initial coupled process state $<X_v,Y_v>$ as follows:
$X_v = R$ if $v \in A_R$, $X_v = B$ if $v \in A_B$, and $X_v = U$ otherwise;
and $Y_v = R$ if $v \in A_R$, and $Y_v = U$ otherwise.
It is easily verified that these initial
states satisfy Properties~\ref{prop1} and~\ref{prop3} above, thus encoding the initial states
of the two separate processes.

Assume for now that the first vertex or vertices $v$ to be updated in the $X$ and $Y$
processes are the same --- i.e. the same vertices are updated in both the joint and solo
update schedules, which may in general depend on the state of the network in each. We
now describe the coupled updates of $v$.
% Now consider the first vertex $v$ to be updated in the coupled process.
Let $\alpha^R_v$
denote the fraction of $v$'s neighbors $w$ such that $X_w = R$, and $\alpha^B_v$
the fraction such that $X_w = B$. Note that by the initialization of the coupled process,
$\alpha^R_v$ is also equal to the fraction of $Y_w = R$ (which we denote
$\tilde{\alpha}^R_v$).

In the joint process, the probability that $v$ is updated to $R$ is
$h(\alpha^R_v,\alpha^B_v)$, and to $B$ is
$h(\alpha^B_v,\alpha^R_v)$.
In the solo Red process, the probability that $v$ is updated to $R$ is
$h(\alpha^R_v,0)$, which by competitiveness is greater than or equal to
$h(\alpha^R_v,\alpha^B_v)$.

\begin{figure}[h]
 \begin{center}
  \includegraphics[scale=.3]{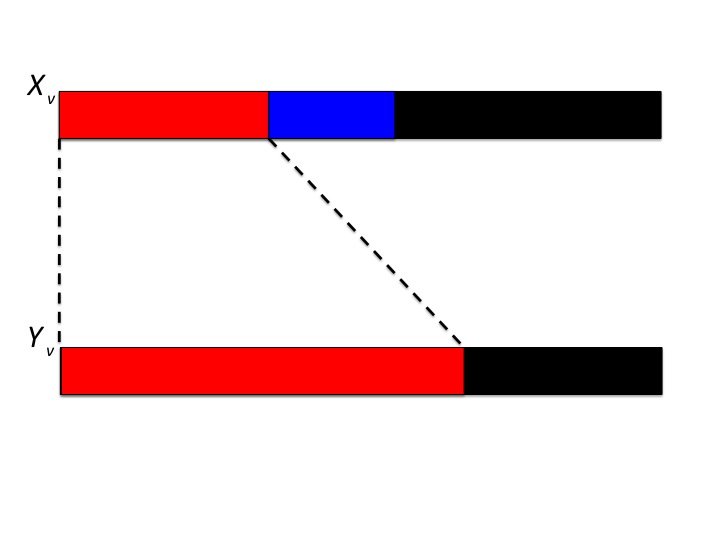}
\end{center}
    \vspace{-0.5in}
\caption{
 Illustration of the coupled dynamics defined in the proof of
 Lemma~\ref{lem:coupled}. In the update dynamic for $X_v$ (top line), the
probabilities of Red and Blue updates are represented by disjoint line
segments of length
$h(\alpha^R_v,\alpha^B_v)$ and
$h(\alpha^B_v,\alpha^R_v)$ respectively.
By competitiveness, the Red segment has length less than
$h(\alpha^R_v,0)$, which is
the probability of Red update of $Y_v$ (bottom line). The dashed red lines
indicate this inequality.
Thus by the arrangement of the line segments we enforce
the invariant that
$X_v = R$ implies $Y_v = R$.
}
  \label{fig:coupled}
\end{figure}

We can thus define the update dynamics of the coupled process as follows: pick a
real value $z$ uniformly at random from $[0,1]$. Update the state $<X_v,Y_v>$ of $v$
as follows:
\begin{itemize}
\item $X_v$ update: If
$z \in [0,h(\alpha^R_v,\alpha^B_v))$, update $X_v$ to $R$; if
$z \in [h(\alpha^R_v,\alpha^B_v),h(\alpha^R_v,\alpha^B_v)+h(\alpha^B_v,\alpha^R_v)]$, update $X_v$ to $B$;
otherwise, update $X_v$ to $U$. Note that the probabilities $X_v$ are updated to $R$ and
$B$ exactly match those of the joint process, as required by Property~\ref{prop2} above.
See Figure~\ref{fig:coupled}.
\item $Y_v$ update: If
$z \in [0,h(\alpha^R_v,0)]$, update $Y_v$ to $R$;
otherwise, update $Y_v$ to $U$.
The probability $Y_v$ is updated to $R$ is thus exactly
$h(\alpha^R_v,0)$, matching that in a solo Red process.
See Figure~\ref{fig:coupled}.
\end{itemize}
Since by competitiveness,
% $r \in [0,h^R(\alpha^R_v,\alpha^B_v)+h^B(\alpha^R_v,\alpha^B_v))$
$z \in [0,h(\alpha^R_v,\alpha^B_v))$
implies
$z \in [0,h^R(\alpha^R_v,0)]$,
we ensure
Property~\ref{prop3}.
Thus in subsequent updates we shall have
$\alpha^R \leq \tilde{\alpha}^R$.
Thus as long as
$h(\alpha^R,\alpha^B) \leq h(\tilde{\alpha}^R,0)$
we can continue to
maintain the invariant. These inequalities
follow from competitiveness and the additivity of $H$.
% and $h$ being increasing in its first argument.

So far we have assumed the same vertices were candidates for updating in both the joint and
solo processes; while this may be true for some update schedules, in general it will not be
(such as in parallel updates of all uninfected vertices, where vertices with only blue
neighbors in the joint process will not be candidates for updating in the red solo process).
However, this is easily handled by considering three cases. Case 1: Assuming
Property~\ref{prop3} holds, if a vertex is a candidate for updating in both processes, we
can maintain this property by performing the coupled updates described above.
Case 2: If a vertex $v$ is a candidate for updating only in the solo red process, then by
Property~\ref{prop3}
$X_v$ cannot be $R$, so
Property~\ref{prop3}
will still hold after the update of $Y_v$.
Case 3: Finally, if $v$ is a candidate for updating only in the joint process, then if $Y_v = R$,
Property~\ref{prop3}
will still hold after the update of $X_v$, and if $Y_v = U$ and all neighbors of $v$ in
the joint process are $B$,
Property~\ref{prop3} will remain true after the update.
The only case remaining is that
$Y_v = U$ and $v$ has $R$ neighbors in the joint process.
This is impossible for the update schedules
mentioned in Section~\ref{sec:contsec}:
$v$ should have also been a candidate for updating in the
solo red process, since by Property~\ref{prop3}
$v$ has weakly more $R$ neighbors in the solo process.

Since Properties~\ref{prop2} and~\ref{prop3} hold on an update-by-update basis in any run or sample
path of the coupled dynamics, they also hold in expectation over runs, yielding the
statement of the lemma.
% Note also that since the lemma holds for every fixed sets $A_R$ and $A_B$,
% it also holds if $A_R$ and $A_B$ are chosen randomly (mixed strategies).
\qed\ (Lemma~\ref{lem:coupled})
\bigskip

\begin{corollary} \label{corr:coupled}
Let $A_R$ and $A_B$ be any sets of seeded nodes for the two players. Then if
the adoption function $h(\alpha^R,\alpha^B)$ is competitive and $H$ is additive,
\[
    \Exp[\chi_R+\chi_B|(A_R,A_B)] \leq \Exp[\chi_{R}|(A_R,\emptyset)] + \Exp[\chi_{B}|(\emptyset,A_B)].
\]
\end{corollary}

\noindent
\proof
Follows from linearity of expectation applied to the left hand side of the inequality, and
two applications of Lemma~\ref{lem:coupled}.
\qed\ (Corollary~\ref{corr:coupled})
\bigskip

Let $(S_R^*,S_B^*)$ be the
maximum joint payoff seed sets. Let $(S_R,S_B)$ be any (pure) Nash equilibrium, with $S_R$ having the larger budget.
Corollary~\ref{corr:coupled}
implies either %one of 
$\Exp[\chi_{R}|(S_R^*,\emptyset)]$ or
$\Exp[\chi_{B}|(\emptyset,S_B^*)]$ is at least as great as
$\Exp[\chi_R+\chi_B|(S_R^*,S_B^*)]/2$; so assume without loss of generality that
$\Exp[\chi_{R}|(S_R^*,\emptyset)]
    \geq \Exp[\chi_R+\chi_B|(S_R^*,S_B^*)]/2$.
% and
% $\Exp[\chi_{R}|(S_R,S_B)]
% \leq \Exp[\chi_R+\chi_B|(S_R^*,S_B^*)]/8$.
Let us now contemplate a unilateral deviation of the Red player from $S_R$ to $S_R^*$, in which
case the strategies are $(S_R^*,S_B)$. In the following lemma we show that
{\em total\/} number of eventual adoptions for the two players is larger than adoptions
accruing to a single player under solo seeding.

\begin{lemma} \label{lem:coupled2}
Let $A_R$ and $A_B$ be any sets of seeded nodes for the two players.
If $H$ is additive,
\[
\Exp[\chi_{R} + \chi_{B}|(A_R,A_B)] \geq
    \Exp[\chi_{R}|(A_R,\emptyset)].
\]
\end{lemma}

\noindent
\proof
We employ a coupling argument similar to that in the proof of
Lemma~\ref{lem:coupled}. We define a stochastic process in which the state of a vertex $v$
is a pair $<X_v, Y_v>$ in which the following properties are obeyed:

\begin{enumerate}
\item \label{prop1a} At each step, and for any vertex state $<X_v,Y_v>$,
    $X_v \in \{R,B,U\}$ and
    $Y_v \in \{R,U\}$.
\item \label{prop2a} Projecting the state of the coupled process onto either component
    faithfully yields the respective process. Thus, if
    $<X_v,Y_v>$ represents the state of vertex $v$ in the coupled process, then
    the $\{X_v\}$ are stochastically identical to the joint process $(A_R,A_B)$,
    and the $\{Y_v\}$ are stochastically identical to the solo Red process $(A_R,\emptyset)$.
\item \label{prop3a} At each step, and for any vertex state $<X_v,Y_v>$,
    $Y_v = R$  implies
    $X_v = R$
    or $X_v = B$.
\end{enumerate}

We initialize the coupled process in the obvious way:
if $v \in A_R$ then $X_v = R$,
if $v \in A_B$ then $X_v = B$,
and $X_v = U$ otherwise; and
if $v \in A_R$ then $Y_v = R$,
and $Y_v = U$ otherwise. Let us fix a vertex $v$ to update, and let
$\alpha^R_v, \alpha^B_v$ denote the fraction of neighbors $w$ of $v$ with
$X_w = R$ and $X_w = B$ respectively, and let
$\tilde{\alpha}^R_v$ denote the fraction with
$Y_w = R$. Initially we have $\alpha^R_v = \tilde{\alpha}^R_v$.

\begin{figure}[h]
  \begin{center}
    \includegraphics[scale=0.3]{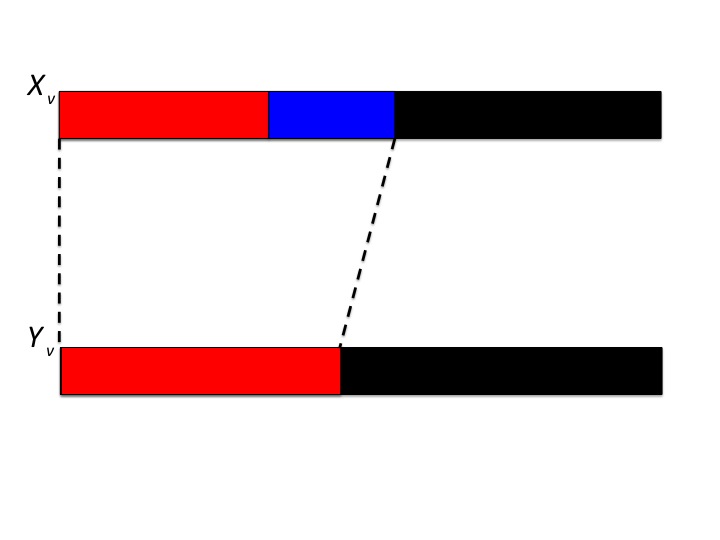}
    \vspace{-0.5in}
 \end{center}
  \caption{
   Illustration of the coupled dynamics defined in the proof of
  Lemma~\ref{lem:coupled2}. In the update dynamic for $X_v$ (top line), the
  probabilities of Red and Blue updates are represented by line
   segments of length
   $h(\alpha^R_v,\alpha^B_v)$ and
  $h(\alpha^B_v,\alpha^R_v)$ respectively.
  By additivity of $H$, together these two segments are greater than
  $h(\alpha^R_v,0)$
   which is the probability of Red update of $Y_v$ (bottom line). This inequality
  is represented by the dashed black lines.
  }
     \label{fig:coupled2}
\end{figure}

We assume the vertex or vertices $v$ to be updated in the $X$ and $Y$
processes are the same; the fact that the update schedules may cause
these sets to differ is handled in the same way as in the proof
of Lemma~\ref{lem:coupled}.
On the first update of $v$ in the joint process $(A_R,A_B)$, the total probability infection by either $R$ or $B$ is
\[
H(\alpha^R_v,\alpha^B_v) =
    h(\alpha^R_v,\alpha^B_v) + h(\alpha^B_v,\alpha^R_v).
\]
In the solo process $(A_R,\emptyset)$, the probability of infection by $R$ is
$h(\alpha^R_v,0) \leq
    h(\alpha^R_v,0) + h(0,\alpha^R_v) = H(\alpha^R_v,0) \leq H(\alpha^R_v,\alpha^B_v)$
where the last inequality follows by the additivity of $H$.

We thus define the update dynamics in the coupled process as follows: pick a real value $z$ uniformly
at random from $[0,1]$. Update $<X_v,Y_v>$ as follows:
\begin{itemize}
\item $X_v$ update: \\
    If $z \in [0,h(\alpha^R_v,\alpha^B_v))$, update $X_v$ to $R$;\\
    if $z \in [h(\alpha^R_v,\alpha^B_v),h(\alpha^R_v,\alpha^B_v) + h(\alpha^B_v,\alpha^R_v)]\\
       \equiv [h(\alpha^R_v,\alpha^B_v),H(\alpha^R_v,\alpha^B_v)]$,
    update $X_v$ to $B$;\\
    otherwise update $X_v$ to $U$.
    See Figure~\ref{fig:coupled2}.
\item $Y_v$ update: If $r \in [0,h(\alpha^R_v,0))$, update $Y_v$ to $R$;\\
    otherwise update $Y_v$ to $U$.
    See Figure~\ref{fig:coupled2}.
\end{itemize}
It is easily verified that at each such update, the probabilities of $R$ and $B$ updates of $X_v$ are
exactly as in the joint $(A_R,A_B)$ process, and the probability of an $R$ update of $Y_v$ is exactly
as in the solo $(A_R,\emptyset)$ process, thus maintaining Property~\ref{prop2a} above.
Property~\ref{prop3a} follows from the previously established fact that
$h(\alpha^R_v,0) \leq H(\alpha^R_v,\alpha^B_v)$, so whenever $Y_v$ is updated to $R$, $X_v$ is updated to
either $R$ or $B$.

Notice that since $h(\alpha^R_v,0) \geq h(\alpha^R_v,\alpha^B_v)$ by competitiveness, for the overall
theorem (which requires competitiveness of  $h$) we {\em cannot\/} ensure that $Y_v = R$
is always accompanied by $X_v = R$. Thus the Red infections in the solo process may exceed those in the joint
process, yielding $\tilde{\alpha}^R_v > \alpha^R_v$ for subsequent updates. To maintain Property~\ref{prop3a} in
subsequent updates we thus require that
$\tilde{\alpha}^R_v \leq
    \alpha^R_v + \alpha^B_v$
implies
$h(\tilde{\alpha}^R_v,0) \leq H(\tilde{\alpha}^R_v,0) \leq H(\alpha^R_v,\alpha^B_v)$
which follows from the additivity of $H$.
Also, notice that since the lemma holds for every fixed $A_R$ and $A_B$, it also holds in expectation for mixed strategies.
\qed\ (Lemma~\ref{lem:coupled2})
\bigskip

Continuing the analysis of a unilateral deviation by the Red player from $S_R$ to $S_R^*$, we have
thus established
\begin{eqnarray*}
\lefteqn{\Exp[\chi_{R} + \chi_{B}|(S_R^*,S_B)]}\\
& = &
    \Exp[\chi_{R}|(S_R^*,S_B)] +
    \Exp[\chi_{B}|(S_R^*,S_B)] \\
    & \geq & \Exp[\chi_{R}|(S_R^*,\emptyset)] \\
    & \geq & \Exp[\chi_R+\chi_B|(S_R^*,S_B^*)]/2
\end{eqnarray*}
where the equality is by linearity of expectation, the first inequality follows from Lemma~\ref{lem:coupled2},
and the second inequality from Corollary~\ref{corr:coupled}.
Thus at least one of
$\Exp[\chi_{R}|(S_R^*,S_B)]$
and $\Exp[\chi_{B}|(S_R^*,S_B)]$ must be at least
$\Exp[\chi_R+\chi_B|(S_R^*,S_B^*)]/4$.

If
$\Exp[\chi_{R}|(S_R^*,S_B)]
    \geq \Exp[\chi_R+\chi_B|(R^*,B^*)]/4$,
then since $(S_R,S_B)$ is Nash, %we must also have
$\Exp[\chi_{R}|(S_R,S_B)]
    \geq \Exp[\chi_R+\chi_B|(S_R^*,S_B^*)]/4$, and the theorem is proved.
%Thus the 
The only remaining case is where
$\Exp[\chi_{B}|(S_R^*,S_B)]
    \geq \Exp[\chi_R+\chi_B|(S_R^*,S_B^*)]/4$.
But Lemma~\ref{lem:coupled} has already established that
$\Exp[\chi_{B}|(\emptyset,S_B)]
    \geq \Exp[\chi_{B}|(S_R^*,S_B)]$, and %from Lemma~\ref{lem:coupled2}
we have
$\Exp[\chi_{R} + \chi_{B}|(S_R,S_B)]
    \geq \Exp[\chi_{B}|(\emptyset,S_B)]$ from Lemma~\ref{lem:coupled2}.
Combining, we have the following chain of inequalities:
\begin{eqnarray*}
\Exp[\chi_{R} + \chi_{B}|(S_R,S_B)]
    & \geq & \Exp[\chi_{B}|(\emptyset,S_B)] \\
    & \geq & \Exp[\chi_{B}|(S_R^*,S_B)] \\
    & \geq & \Exp[\chi_R+\chi_B|(S_R^*,S_B^*)]/4
\end{eqnarray*}
thus establishing the theorem.
\qed\ (Theorem~\ref{theoremPOA1})
\bigskip

{\bf Concave $f$, non-linear $g$.\/}
Recall that the switching-selection formulation in which $f$ is concave and $g$ is
linear satisfies the hypothesis of the Theorem above.
But Theorem~\ref{theoremPOA1} also provides more general conditions for bounded
PoA in the switching-selection model.
For example, suppose we consider switching functions of the form
$f(x) = x^r$ for $r \leq 1$ (thus yielding concavity) and selection
functions of the Tullock contest form
$g(y) = y^s/(y^s + (1-y)^s)$, as discussed in Section~\ref{sec:contsec}.
Letting $a$ and $b$ denote the local fraction of Red and Blue neighbors for
notational convenience, this leads to an adoption function of the form
$h(a,b) = (a+b)^r/(1 + (b/a)^s)$. The condition for competitiveness is
\[
h(a,0) - h(a,b) =
    a^r - (a+b)^r/(1 + (b/a)^s) \geq 0.
\]
Dividing through by $(a+b)^r$ yields
\begin{eqnarray*}
\lefteqn{(a/(a+b))^r - 1/(1 + (b/a)^s)} \\
& & = 1/(1 + (b/a))^r - 1/(1 + (b/a)^s) \geq 0.
\end{eqnarray*}
Making the substitution $z = b/a$ and moving the second term to the right-hand side gives
\[
1/(1+z)^r \geq 1/(1+z^s).
\]
Thus competitiveness is equivalent to the condition $1 + z^s \geq (1+z)^r$ for all $z \geq 0$.
\iffalse
Consider the choice $r = 1/2$. In this case the condition becomes
$1 + z^s \geq \sqrt{1+z}$, or
$(1+z^s)^2 \geq 1 + z$. Expanding gives
$1 + 2z^s + z^{2s} \geq 1+z$. It is easily seen this inequality is obeyed for all
$z \geq 0$ provided $s \in [1/2,1]$, since then for all $z \geq 1$ we have $z^{2s} \geq z$,
and for all $z \leq 1$ we have $z^s \geq z$. More generally, if we let $r = 1/k$ for
some natural number $k$, the competitiveness condition becomes $(1+z^s)^k \geq 1+z$.
The smallest power of $z$ generated by the left-hand side is $z^s$ and the largest is $z^{sk}$.
As long as $sk \geq 1$, $z^{sk} \geq z$ for $z \geq 1$, and as long
as $s \leq 1$ then $z^s \geq z$ for $z \leq 1$. 
\fi
It is not difficult to show that any $s \in [r,1]$ will satisfy this condition.
In other words, the more concave $f$ is (i.e. the smaller $r$ is),
the more equalizing $g$ can be (i.e. the smaller $s$ can be) while maintaining
competitiveness. By Theorem~\ref{theoremPOA1} we have thus shown:

\begin{corollary}
Let the switching function be $f(x) = x^r$ for $r \leq 1$ and %the selection function be
let $g(y) = y^s/(y^s + (1-y)^s)$ be the selection function.
Then as long as $s \in [r,1]$, the Price of Anarchy is at most 4 for any graph.% $G$.
\end{corollary}

\subsection{PoA: Lower Bound}

We now show that concavity of the switching function is required in a
very strong sense --- essentially, even
a slight convexity leads to unbounded PoA. As a first step in this
demonstration, it is useful to begin with a simpler result showing that the PoA is unbounded if the
switching function is permitted to violate concavity to an {\em arbitrary\/} extent.

\begin{theorem} \label{theoremPOA2}
Fix $\alpha^* \in (0,1)$, and let the switching function $f$ be the threshold function
$f(x) = 0$ for $x < \alpha^*$, and $f(x) = 1$ for $x \geq \alpha^*$. Let the selection
function be linear $g(y) = y$. Then for any value $V > 0$, there exists $G$ such
that the Price of Anarchy in $G$ is greater than $V$.
\end{theorem}

\noindent
\proof
Let $m$ be a large integer, and set the initial budgets of both players to
be $\alpha^* m/2$. The graph $G$ will consist of two components. The first
component $C_1$ consists of two layers; the first layer has $m$ vertices and the
second $n_1$ vertices, and there is a directed edge from every vertex in the
first layer to every vertex in the second layer. The second component $C_2$ has
the same structure, but with $m$ vertices in the first layer and $n_2$
in the second layer. We let $n_2 \gg n_1 \gg m$. For concreteness, let us
choose an update schedule that updates each vertex in the second layers of
the two components exactly once in some fixed ordering (the same result
holds for many other updating schedules).

It is easy to see that the maximum joint profit solution is to place the
combined $\alpha^* m$ of seeds of the two players in the first
layer $C_2$, in which case the number of second-layer infections
will be $n_2$ since $f(\alpha^*) = 1$. Any configuration which places at least
one infection in each of the two components will not cause any second-layer
infections, since then the threshold of $f$ will not be exceeding in either
component.

It is also easy to see that both players placing all their infections in
the first layer of $C_1$, which will result in $n_1$ infections in the
second layer since the threshold is exceeded, is a Nash equilibrium.
Any deviation of a player to $C_2$, or to layer 2 of $C_1$, causes the
threshold to no longer be exceeded in either component. Thus the PoA here
is $n_2/n_1$, which can be made arbitrarily large. Note that the
maximum joint infections solution is also a Nash equilibrium --- we are
exploiting the worst-case (over Nash) nature of the PoA here (as will
all our lower bounds, though see Footnote~\ref{hhnote}).
\qed\ (Theorem~\ref{theoremPOA2})
\bigskip

Thus, a switching function strongly violating concavity can lead to
unbounded PoA even with a linear selection function.
But it turns out that functions even {\em slightly\/} violating concavity
also cause unbounded PoA --- as we shall see, network structure can
{\em amplify\/} small amounts of
convexity.\footnote{The theorem which follows considers the family $f(\alpha)=\alpha^r$, but can be 
generalized to other choices of convex $f$ as well.}

\begin{theorem}\label{theoremPOA3}
Let the switching function be $f(x) = x^r$ for any $r > 1$, and let
the selection function be linear $g(y) = y$. Then for any $V > 0$, there exists a
graph $G$ for which the Price of Anarchy is greater than $V$.
More precisely, there is a family of graphs for which the Price of Anarchy grows
linearly with the population size (number of vertices).
\end{theorem}

\begin{figure*}[tbh]
 \begin{center}
 \includegraphics[scale=.23]{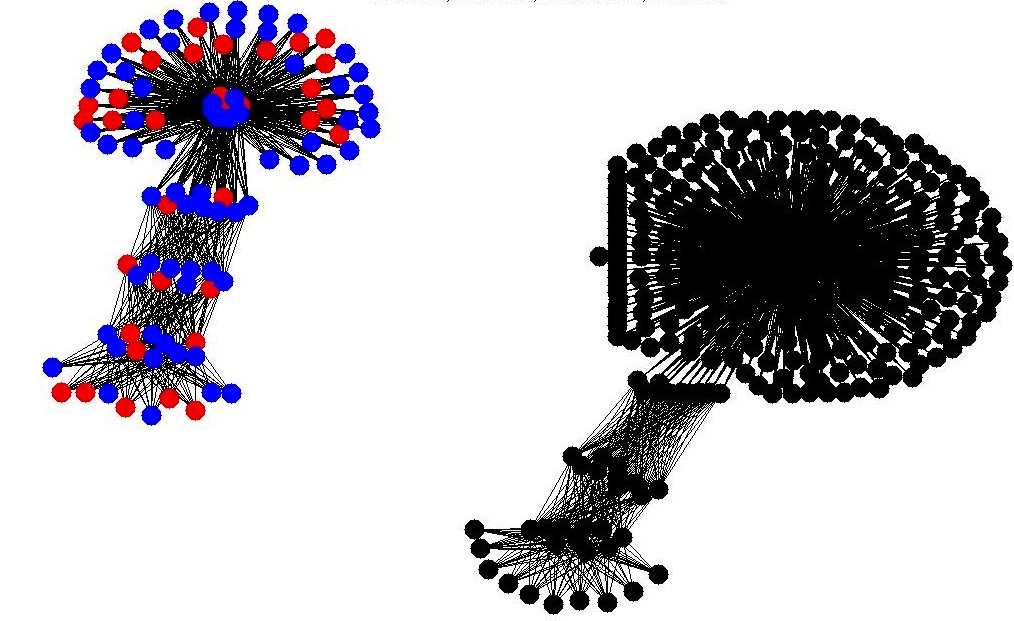}
 \newline
    \includegraphics[scale=.23]{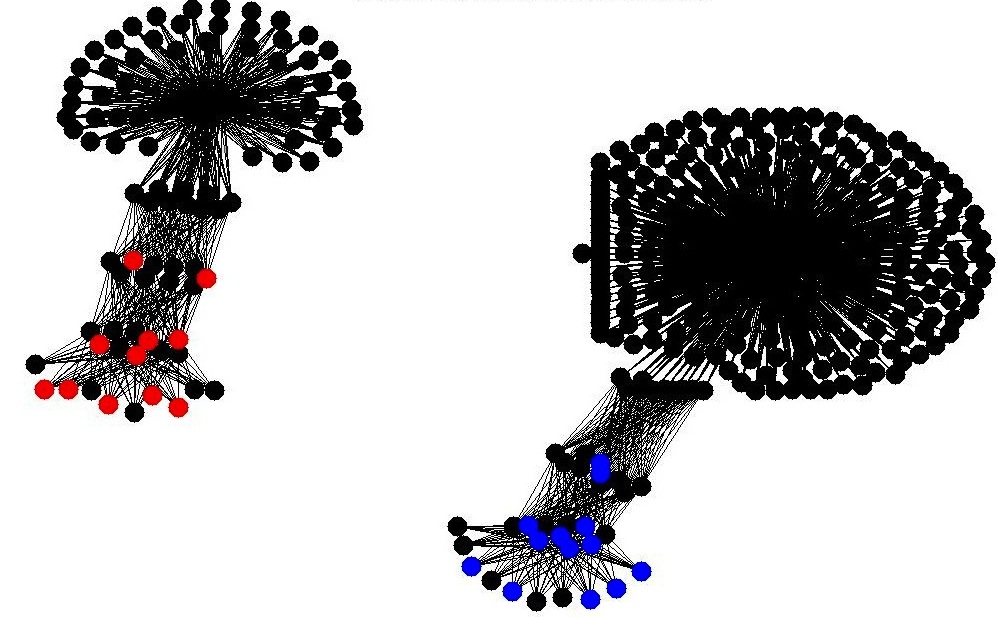}
  \end{center}
% \vspace{-0.5in}
  \caption{
    Illustration of convexity amplification in the Price of Anarchy lower bound of
    Theorem~\ref{theoremPOA3}, under
    convex switching function $f(x) = x^3$ and linear selection function $g$.
    Top: Two-component, directed, layered ``flower'' graph, with the right flower having many more
    petals than the left. In the configuration shown, both players play in the first layer of the stem
    of the smaller flower. The convexity of $f$ does not enter the dynamics, since at each update
    an entire successive layer is infected, quickly reaching the petals.
    % There is also a better equilibrium (not shown), in which the players play in the first layer of the larger flower,
    % and again there is complete infection, but with larger payoffs.
    Bottom: However, if the two players locate in different components, layers are not
    fully infected and the convexity of $f$ is amplified via composition in successive layers, damping out
    the infection rate quickly.
    }
        \label{fig:PoAamp}
\end{figure*}

\noindent
\proof
The idea is to create a layered, directed graph whose dynamics rapidly amplify the convexity of $f$.
Taking two such {\em amplification components\/} of differing sizes
yields an equilibrium in which the players coordinate on
the smaller component, while the maximum joint payoffs solution lies in the larger component.
The construction of the proof is illustrated in Figure~\ref{fig:PoAamp}.

The amplification gadget
will be a layered, directed graph with $\ell_i$ vertices in the $i$th layer and $N$ layers total.
There are directed edges from every vertex in layer $i$ to every vertex in layer $i+1$, and
no other edges.
Let the two players have equal budgets of $k$, and define
$\alpha = 2k/\ell_1$ --- thus, $\alpha$ is the fraction of layer 1 the two players could
jointly infect.

Let us consider what happens if indeed the two players jointly infect $2k$ vertices in the
first layer, and the update schedule proceeds by updating each successive layer $2,\ldots,N$.
Since every vertex in layer 2 has every vertex in layer 1 as a directed neighbor, and no others,
the expected fraction of layer 2 that is infected is $f(\alpha) = \alpha^r$. Inductively, the expected fraction
of layer 3 that is infected is thus $f(f(\alpha)) = \alpha^{r^2}$. In general, the expected
fraction of layer $i$ that is infected is $\alpha^{r^{i-1}}$, and by the linearity of $g$ the
two players will split these infections. Here, we note that the actual path of infections will be stochastic;
this stochastic path is well approximated by the expected infections, if layers are sufficiently large.  Throughout
this proof we will use this approximation (which relies on an appeal to the strong law of large numbers).

Now let $\alpha = \beta_1 + \beta_2$, and let us
instead place
$\beta_1\ell_1$ seeds at layer 1 and
$\beta_2\ell_1$ at layer $i$.
The total number of infections expected at layer $i$ now becomes
$\beta_1^{r^{i-1}}\ell_i + \beta_2\ell_1$. By the convexity of the function
$f^{(i-1)}(x) = x^{r^{i-1}}$, this will be smaller than
$\alpha^{r^{i-1}}\ell_i =
    (\beta_1+\beta_2)^{r^{i-1}}\ell_i$ as long as
$\beta_2\ell_1 < \beta_2^{r^{i-1}}\ell_i$, or $\ell_i > \ell_1/\beta_2^{r^{i-2}}$.
Also, notice that the smallest nonzero deviation requires
$\beta_2\ell_1 \geq 1$, or $\beta_2 \geq 1/\ell_1$. Thus as long as
$\ell_i \geq \ell_1^{r^{i-1}}$, the total fraction of infections generated by
placing
$\beta_1\ell_1$ seeds at layer 1 and
$\beta_2\ell_1$ at layer $i$ will be less than by placing all in layer 1.
Furthermore, by the linearity of $g$, any individual player who effects such a unilateral deviation
will suffer.

Note that we can make the final, $N$th, layer arbitrarily large. In particular,
if we choose
$\ell_i = \ell_1^{r^{i-1}}$ as specified above for all $2 \leq i \leq N-1$, and
choose $\alpha^{r^{N-1}}\ell_N \gg \sum_{i=1}^{N-1} \alpha^{r^{i-1}}\ell_i$,
the total expected number of infections conditioned
on both players playing in the first layer will be dominated by
the $\alpha^{r^{N-1}}\ell_N$
expected infections in the final layer.

Now consider a graph consisting of {\em two \/} disjoint amplification gadgets $G_1$ and $G_2$
that are exactly as described above, but differ only in the sizes of their final $N$th layers ---
$\ell_N(1)$ for $G_1$
and $\ell_N(2)$ for $G_2$, where we will choose $\ell_N(2) \gg \ell_N(1)$.
Consider a configuration where all seeds are in the first layer of $G_1$. We have
already argued above that no deviation to later layers of $G_1$ can be profitable. Now let us
consider a unilateral deviation of the Red player from $G_1$ to the first layer of $G_2$. Since
Red alone infects now only infects a fraction $\alpha/2$ of the $\ell_1$ vertices in the first
layer of $G_2$, the expected final number of Red infections will be approximately
$(\alpha/2)^{r^{N-1}}\ell_N(2)$, compared with
$\alpha^{r^{N-1}}\ell_N(1)/2$ for not deviating from $G_1$. Thus as long as
$(\alpha/2)^{r^{N-1}}\ell_N(2) \leq
    \alpha^{r^{N-1}}\ell_N(1)/2$, or
$\ell_N(2)/\ell_N(1) \leq 2^{r^{N-1}-1}$, this deviation is unprofitable for Red.
More generally, if Red divides its
$(\alpha/2)\ell_1$
resources by placing a fraction $\beta_1$ of them in the first layer of $G_1$ and
a fraction $\beta_2 = 1 - \beta$ of them in the first layer of $G_2$, its expected payoff is
\[
\left[ (1+\beta_1)(\alpha/2) \right] ^{r^{N-1}} \ell_N(1) \frac{\beta_1}{1+\beta_1}
    + \left[ \beta_2(\alpha/2) \right]^{r^{N-1}} \ell_N(2).
\]
The first term of this sum represents the share of the final layer of $G_1$ that
Red obtains given that Blue is playing entirely in this component, while the second
term represents the uncontested infections Red wins in $G_2$.
This expression can be
written as
\[
\alpha^{r^{N-1}} \left[
    \left( \frac{1+\beta_1}{2} \right)^{r^{N-1}} \ell_N(1) \frac{\beta_1}{1+\beta_1}
    + \left( \frac{1-\beta_1}{2}\right)^{r^{N-1}} \ell_N(2)
\right]
\]
which for the choice $\ell_N(2) = 2^{r^{N-1}} \ell_N(1)/2$ becomes
\[
\alpha^{r^{N-1}} \left[
    \left( \frac{1+\beta_1}{2} \right)^{r^{N-1}} \ell_N(1) \frac{\beta_1}{1+\beta_1}
    + \left( {1-\beta_1} \right)^{r^{N-1}} \ell_N(1)/2
\right].
\]
For any $0 < \beta_1 < 1$, both terms inside the brackets above are
exponentially damped and result in suboptimal payoff for Red. Thus the best
response choices are given by the extremes $\beta_1 = 1$ and $\beta_1 = 0$, which
both yield expected payoff $\ell_N(1)/2$ for Red. (Note that by choosing $\ell_N(2)$
slightly smaller above, we can force $\beta_1 = 1$ to be a strict best response.)

However, the maximum joint payoffs solution (as well as the best, as opposed to worst
Nash equilibrium) is for {\em both\/} players to initially infect in the first layer of $G_2$,
in which case the total payoff will be approximately $\alpha^{r^{N-1}}\ell_N(2)$. The
Price of Anarchy is thus
\[
    \frac{\alpha^{r^{N-1}}\ell_N(2)}
        {\alpha^{r^{N-1}}\ell_N(1)} =
    \frac{\ell_N(2)}{\ell_N(1)} \geq
    2^{r^{N-1}-1}
\]
by the choice of $\ell_N(2)$ above.
Thus by choosing the number of layers $N$ as large as needed, the Price of Anarchy
exceeds any finite bound $V$.
\qed\ (Theorem~\ref{theoremPOA3})
\bigskip
\newpage
Combining Theorem~\ref{theoremPOA1} and Theorem \ref{theoremPOA3}, we note that for
$f(x) = x^r$ and linear $g$ we obtain the following sharp threshold result:

\begin{corollary} \label{cor:PoAthold}
Let the switching function be $f(x) = x^r$, and let the
selection function be linear, $g(y) = y$. Then:
\begin{itemize}
\item For any $r \leq 1$, the Price of Anarchy is at most 4 for any graph $G$;
\item For any $r > 1$ and any $V$, there exists a graph $G$ for which the Price of Anarchy
    is greater than $V$.
\end{itemize}
\end{corollary}

\section{Results: Budget Multiplier}\label{sectionPOB}

We derive sufficient conditions for bounded Budget Multiplier, and show that violations of these
conditions can lead to unbounded Budget Multiplier.

\subsection{Budget Multiplier: Upper Bound}

As in the PoA analysis, it will be technically convenient to return to the generalized adoption function model.
Recall that for PoA, competitiveness of $h$ and additivity of $H$ were
needed to prove upper bounds, but we didn't require that the implied
selection function be linear. Here we introduce that additional
requirement, and prove that the (pure strategy) Budget Multiplier is bounded.

% \mk{What and where should we say about the mixed case? Probably need to mention the
% restriction to pure back in the Intro.}

\begin{theorem}\label{theoremPOB1}
Suppose the adoption functions $h(\alpha^R,\alpha^B)$
is competitive, that $H$ is additive in its arguments, and that the 
implied selection function is linear:
\[g(\alpha_R,\alpha_B) = \frac{h(\alpha_R,\alpha_B)} 
{h(\alpha_R,\alpha_B) + h(\alpha_B,\alpha_B)} = \alpha_R/(\alpha_R + \alpha_B)\]
Then the pure strategy Budget Multiplier is at 
most 2 for any graph $G$.\footnote{The theorem actually holds for any equilibrium in which the player with the larger
budget plays a pure strategy; the player with smaller budget may always play mixed. It is easy
to find cases with such equilibria. The theorem also holds for general mixed strategies under
certain conditions --- for instance, when both $f$ and $g$ are linear and the larger budget
is an integer multiple of the smaller.}
\end{theorem}

\noindent
\proof
The proof borrows elements from the proof of Theorem~\ref{theoremPOA1},
and introduces the additional notion of tracking or attributing indirect
infections generated by the dynamics to specific seeds.

Consider any pure Nash equilibrium given by seed sets $S_R$
and $S_B$ in which $|S_R| = K > |S_B| = L$.
For our purposes the interesting case is
one in which
\[
\Exp[\chi_{R}|(S_R,S_B)] \geq \Exp[\chi_{B}|(S_R,S_B))]
\]
and so
\[
\Exp[\chi_{R}|(S_R,S_B)] \geq \Exp[\chi_R+\chi_B|(S_R,S_B)]/2.
\]
Since the
adoption function is competitive and additive, Lemma~\ref{lem:coupled} implies that
$\Exp[\chi_{R}|(S_R,\emptyset)] \geq E[\chi_R|(S_R,S_B)]$ --- that is, the Red player
only benefits from the departure of the Blue player.

Let us consider the dynamics of the solo Red process given by $(S_R,\emptyset)$.
We first introduce a faithful simulation of these dynamics that also allows us to
attribute subsequent infections to exactly one of the seeds in $S_R$;
we shall call this process the {\em attribution simulation\/} of $(S_R,\emptyset)$.
Thus, let $S_R = \{v_1,\ldots,v_K\}$ be the initial Red infections, and let us
label $v_i$ by $R_i$,
and label all other vertices $U$.
All infections in the process will also be assigned one of
the $K$ labels $R_i$ in the following manner: when updating a vertex $v$, we first
compute the fraction $\alpha^R_v$ of neighbors whose current label is one of
$R_1,\ldots,R_K$, and with probability $H(\alpha^R_v,0) = h(\alpha^R_v,0) + h(0,\alpha^R_v)$
we decide that an infection
will occur (otherwise the label of $v$ is updated to $U$). If an infection occurs,
we simply choose an infected neighbor of $v$ uniformly at random, and update $v$
to have the same label (which will be one of the $R_i$). It is easily seen that
at every step, the dynamics of the $(S_R,\emptyset)$ process are faithfully
implemented if we drop label subscripts and simply view any label $R_i$ as a
generic Red infection $R$. Furthermore, at all times every
infected vertex has only one of the labels $R_i$. Thus if we denote the
expected number of vertices with label $R_i$ by
$\Exp[\chi_{R_i}|(S_R,\emptyset)]$, we have
$\Exp[\chi_R|(S_R,\emptyset)] =
    \sum_{i=1}^K \Exp[\chi_{R_i}|(S_R,\emptyset)]$.
Let us assume without loss of generality that the labels $R_i$ are sorted in order of decreasing
$\Exp[\chi_{R_i}|(S_R,\emptyset)]$.

We now consider the payoff to Blue under a deviation from $S_B$ to
the set $\hat{S}_B = \{v_1,\ldots,v_L\} \subset S_R$ --- that is, the $L$ ``most profitable'' initial
infections in $S_R$.
Our goal is to show that the
Blue player must enjoy roughly the same payoff from these $L$ seeds as the Red player
did in the solo attribution simulation.

\begin{lemma} \label{lem:coupled3}
\begin{eqnarray*} \Exp[\chi_B|(S_R,\hat{S}_B)] & \geq &
    \frac{1}{2}\sum_{i=1}^L \Exp[\chi_{R_i}|(S_R,\emptyset)] \\
    & \geq & \frac{L}{2K}\Exp[\chi_R|(S_R,\emptyset)]
\end{eqnarray*}
\end{lemma}

\noindent
\proof
The second inequality follows from
\[
\Exp[\chi_R|(S_R,\emptyset)] =
    \sum_{i=1}^K \Exp[\chi_{R_i}|(S_R,\emptyset)],
\]
established above, and
fact that the vertices in $S_R$ are ordered in decreasing profitability.
For the first inequality, we introduce coupled attribution simulations for the
two processes $(S_R,\emptyset)$ (the solo Red process) and $(S_R,\hat{S}_B)$.
For simplicity, let us actually
examine
$(S_R,\emptyset)$ and $(S_R-\hat{S}_B,\hat{S}_B)$; the latter joint process is simply
the process $(S_R,\hat{S}_B)$, but in which the contested seeded nodes in
$\hat{S}_B$ are all won by the Blue player. (The proof for the general
$(S_R,\hat{S}_B)$ case is the same but causes the factor of $1/2$ in the lemma.)

The coupled attribution dynamics are as follows: as above, in the solo Red process,
for $1 \leq i \leq L$, the vertex $v_i$ in $S_R$ is initially labeled $R_i$,
and all other vertices are labeled $U$.
In the joint process,
the vertex $v_i$ is labeled
$B_i$ for $i \leq L$ (corresponding to the Blue invasions of $S_R$),
while for $L < i \leq K$ the vertex $v_i$ is labeled $R_i$ as before.
Now at the first update vertex $v$, let
$\alpha^R_v$
be the fraction of
Red neighbors in the solo process, and let
$\tilde{\alpha}^R_v$
and
$\tilde{\alpha}^B_v$
be the fraction of Red and Blue neighbors, respectively, in the joint process.

Note that initially we have
$\alpha^R_v =
    \tilde{\alpha}^R_v +
    \tilde{\alpha}^B_v$.
Thus by additivity $H$, the total probabilities of infection
$H(\alpha^R_v,0)$ and
$H(\tilde{\alpha}^R_v,\tilde{\alpha}^B_v)$ in the two processes must be identical.
We thus flip a common coin with this shared infection probability to determine whether
infections occur in the coupled process. If not, $v$ is updated to $U$ in both processes.
If so, we now use a coupled attribution step in which we pick an infected neighbor
of $v$ at random and copy its label to $v$ in {\em both \/} processes. Thus if
a label with index $i \leq L$ is chosen, $v$ will be updated to $R_i$ in the
solo process, and to $B_i$ in the
joint process; whereas if $L < i \leq K$ is chosen, the update
will be to $R_i$ in both processes. It is easily verified that each of the two
processes faithfully implement the dynamics of the solo and joint attribution processes, respectively.

This coupled update dynamic maintains two invariants: infections are always matched in the two
processes, thus maintaining
$\alpha^R_v =
    \tilde{\alpha}^R_v +
    \tilde{\alpha}^B_v$
for all $v$ and every step; and for all $i \leq L$, every $R_i$ attribution in the
solo Red process is matched by a $B_i$ attribution in the joint process, thus
establishing the lemma.
\qed\ (Lemma~\ref{lem:coupled3})
\bigskip

Thus, by simply imitating the strategy of the Red player in the $L$
most profitable resources, the Blue player can expect to infect $(1/2)(L/K)$
proportion of infections accruing to Red in isolation.
Since $(S_R,S_B)$ is an equilibrium, the payoffs of Blue
in equilibrium must also respect this inequality.
\qed\ (Theorem~\ref{theoremPOB1})
\bigskip

\subsection{Budget Multiplier: Lower Bound}

We have already seen that concavity of $f$ and linearity of $g$
lead to bounded PoA and Budget Multiplier, and that even slight deviations from
concavity can lead to unbounded PoA. We now show that fixing $f$
to be linear (which is concave), slight deviations from linearity
of $g$ towards polarizing $g$ can lead to unbounded Budget Multiplier, for
similar reasons as in the PoA case: graph structure can
amplify a slightly polarizing $g$ towards arbitrarily high punishment
of the minority player.

\begin{theorem}\label{theoremPOB2}
Let the switching function be $f(x) = x$, and
let the selection function be of Tullock contest form,
$g(y) = y^s/(y^s + ((1-y)^s)$, where $s > 1$.
Then for any $V > 0$, there exists a graph $G$ for which
the Budget Multiplier is greater than $V$.
More precisely, there is a family of graphs for which the Budget Multiplier grows
linearly with the population size (number of vertices).
\end{theorem}
\noindent
\proof
As in the PoA lower bound, the proof relies on a layered amplification graph,
this time amplifying punishment in the selection function rather than convexity
in the switching function.
The graph will consist of two components, $C_1$ and $C_2$.

Let us fix the budget of the Red player to be 3, and that of the Blue
player to be 1 (the proof generalizes to other unequal values).
$C_1$ is a directed, layered graph with $k+1$ layers.
The first layer has $4$ vertices, and layers 2 through $k$ have
$n \gg 4$ vertices, while layer $k+1$ has $n_1$
vertices, where we shall choose $n_1  \gg n$, meaning
that payoffs in $C_1$ are dominated by infections
in the final layer.

The second component $C_2$ is a 2-layer directed graph, with 1 vertex in the
first layer and $n_2$ in the final layer, and all directed edges from layer 1 to 2.
We will eventually choose $n_2 \ll n_1$, so that $C_1$ is the much bigger component.
We choose
an update rule in which each layer is updated in succession and only once.

Consider the configuration in which Red places its 3 infections in the first layer of
$C_1$, and Blue places its 1 infection in the first layer of $C_2$. We shall later show that this configuration is
a Nash equilibrium. In this configuration, the expected payoff to Red is
approximately
$\sum_{i=2}^{k} (3/4)n + (3/4)n_1$
by linearity of $f$; notice that the selection function does not enter since the
players are in disjoint components.
Similarly, the expected payoff to Blue is
$n_2$.
In this configuration, the ratio of Red and Blue expected payoffs is thus at least
$(3/4)n_1/n_2$, whereas the initial budget ratio is $1/3$. So the Budget Multiplier for this
configuration is at least $n_1/(4n_2)$.

We now develop conditions under which this configuration is an equilibrium. It is
easy to verify that red is playing a best response. Moving vertices to later layers of
$C_1$ lowers Red's payoff, since $n \gg 4$ and $f$ is linear. Finally,
moving infections to invade the first layer of $C_2$ will lower Red's payoff as long
as, say,  $(1/4)n_{1}$ (Red's current payoff per initial infection in the final layer of $C_1$) exceeds
$n_2$ (the maximum amount Red could get in $C_2$ by full deviation), or
$n_{1} \gg 4 n_2$.

We now turn to deviations by Blue. Moving the solo Blue initial infection to the
second layer of $C_2$ is clearly a losing proposition.
So consider
deviations in which Blue moves to vertices in component 1. If
he moves to the lone unoccupied vertex in layer 1 of $C_1$, his
payoff is approximately:
\begin{eqnarray*}
\lefteqn{\sum_{i=2}^{k} g^{(i)}(1/4) n
    + g^{(k+1)}(1/4) n_1} & & \\
& = & \sum_{i=2}^{k} \frac{(1/4)^{s^i}}{(1/4)^{s^i} + (3/4)^{s^i}} n\\
& & + \frac{(1/4)^{s^{k+1}}}{(1/4)^{s^{k+1}} + (3/4)^{s^{k+1}}} n_1
\end{eqnarray*}
Similarly, if Blue directly invades a Red vertex, Blue's payoff is approximately
\[
\chi = \sum_{i=2}^{k} \frac{(1/3)^{s^i}}{(1/3)^{s^i} + (2/3)^{s^i}} n
    + \frac{(1/3)^{s^{k+1}}}{(1/3)^{s^{k+1}} + (2/3)^{s^{k+1}}} n_1
\]
Since in both cases Blue's payoff is being exponentially dampened at each successive
layer, it is easy to see that the second deviation is more profitable. Finally, Blue may
choose a vertex in a later layer of $C_1$, but again by $n \gg 4$ and the linearity of $f$,
this will be suboptimal.

Thus as long as we arrange that $n_2$ --- Blue's payoff without deviation --- exceeds
$\chi$ above, we will have ensured that no player has an incentive to deviate from the specified strategy configuration.
Let us scale $n_1$ up as large as necessary to have $\chi$ dominated by the term
involving $n_1$, and now set $n_2$ to equal that term:
\[
n_2  = \frac{(1/3)^{s^{k+1}}}{(1/3)^{s^{k+1}} + (2/3)^{s^{k+1}}} n_1
\]
in order to satisfy the equilibrium condition. The ratio $n_1/n_2$, which we have
already shown above lower bounds the Budget Multiplier, is thus a function that is increasing
exponentially in $k$ for any fixed $s > 1$. Thus by choosing $k$ sufficiently large,
we can force the Budget Multiplier larger than any chosen value.
\qed\ (Theorem~\ref{theoremPOB2})

Combining Theorem~\ref{theoremPOB1} and Theorem \ref{theoremPOB2}, we note that for linear $f$
and Tullock $g$, we obtain the following sharp threshold result, which is analogous to the
PoA result in Corollary~\ref{cor:PoAthold}.

\begin{corollary} \label{cor:PoBthold}
Let the switching function $f$ be linear, and let the selection function $g$ be Tullock,
$g(y) = y^s/(y^s + (1-y)^s)$.  Then:
\begin{itemize}
\item For $s = 1$, the Budget Multiplier is at most 2 for any graph $G$;
\item For any $s > 1$ and any $V$, there exists a graph $G$ for which the Budget Multiplier
    is greater than $V$.
\end{itemize}
\end{corollary}

% \mk{Can/should we add something about the $s < 1$ case above? SG; GOOD POINT.}
In fact, if we permit a slight generalization of our model, in which certain vertices in
the graph are ``hard-wired'' to adopt only one or the other color (so there is no use for
the opposing player to seed them), unbounded Budget Multiplier also holds in the Tullock case for $s < 1$
(equalizing). So in this generalization, linearity of $g$ is required for bounded Budget Multiplier.

\begin{figure*}[tbh]
  \begin{center}
    \includegraphics[scale=.3]{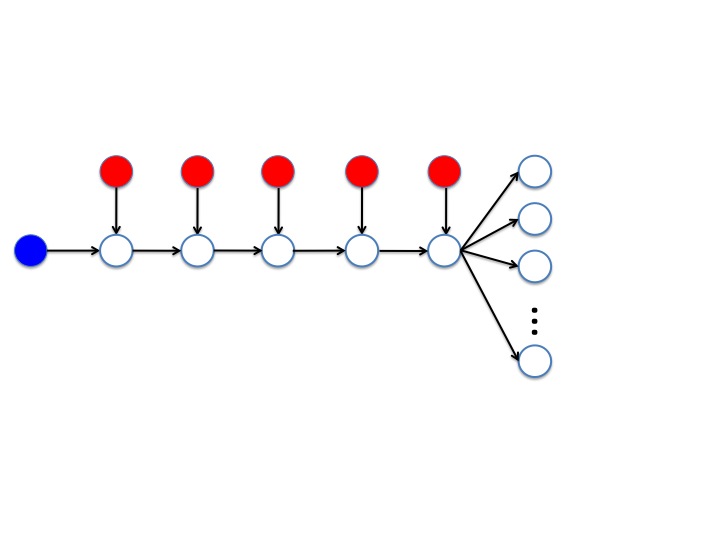}
    \includegraphics[scale=.3]{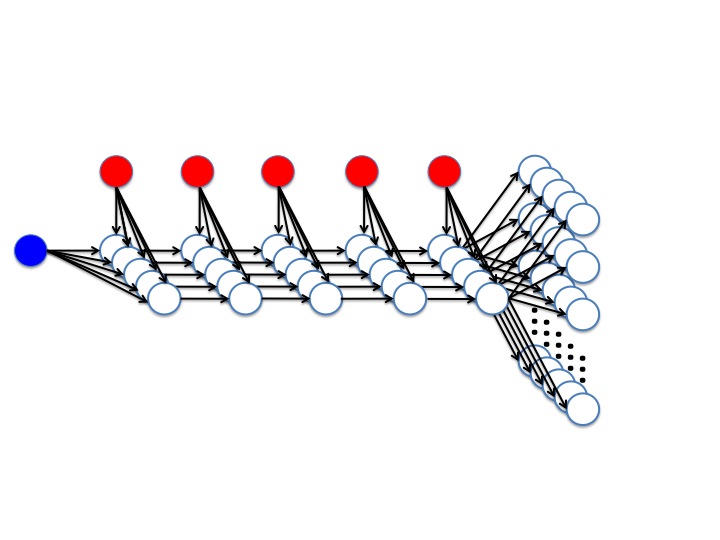}
    \vspace{-0.5in}
  \end{center}
  \caption{Illustration of the construction in the proof of
	Theorem~\ref{theoremPOB3}.
	Left: Basic gadget.
	Right: Equilibrium construction.
    }
\label{fig:tree}
\end{figure*}

We have thus shown that even when the switching function is ``nice'' (linear), even
slight punishment in the selection function can lead to unbounded Budget Multiplier. Recall that
we require switching and selection functions to be 0 (1, respectively) on input
0 (respectively) and increasing, and additionally that $g(1/2) = 1$. The following
theorem shows that if $f$ is allowed to be a sufficiently convex function, then the
Budget Multiplier is again unbounded for {\em any\/} selection function. This establishes the importance
of concavity of $f$ for both the PoA and Budget Multiplier.

\begin{theorem}\label{theoremPOB3}
Let the switching $f$ satisfy $f(1/2) = 0$ and $f(1) = 1$.
Then for any value $V > 0$, there exists $G$ such that the
Budget Multiplier is greater than $V$.
\end{theorem}

\noindent
\proof
Let the Blue player have 1 initial infection and the Red player have $K \geq 2$
(the proof can be generalized to any unequal initial budgets, which we comment on below).
Consider the directed graph shown in the left panel of Figure~\ref{fig:tree},
where we have arranged the 1 Blue and $K$ Red seeded nodes in a particular
configuration. Aside from the initially infected vertices, this graph consists
of a directed chain of $K$ vertices, whose final vertex then connects to a large
number $N \gg K$ of terminal vertices. Let us update each vertex in the chain
from left to right, followed by the terminal vertices.

Let us first compute the expected payoffs for the two players in this configuration.
First, note that since $f(1) = 1$, it is certain that every vertex in the chain will
be infected in sequence, followed by all of the terminal vertices; the only question
is which player will win the most. By choosing $N \gg K$ we can ignore the infections
in the chain and just focus on the terminal vertices, which will be won by
whichever player infects the final chain vertex. It is easy to see that the probability
this vertex is won by Blue is $1/2^K$, since Blue must ``beat'' a competing Red
infection at every vertex in the chain. Thus the expected payoffs are approximately
$N/2^K$ for Blue and $N(1-1/2^K)$ for Red. If this configuration were an equilibrium,
the Budget Multiplier would thus be $2^K/K$, which can be made as large as desired by choosing $K$
large enough.

However, this configuration is not an equilibrium --- clearly,
either player would be better off by simplifying initially infecting
the final vertex of the chain, thus winning all the terminal
vertices. This is fixed by the construction shown in the right panel
of Figure~\ref{fig:tree}, where we have replicated the chain and
terminal vertices $M$ times, but have only the original $K+1$
seeded nodes as common ``inputs'' to all of these
replications. Notice that now if either player defects to an
uninfected vertex, neither player will receive {\em any\/}
infections in any of the other replications, since now there is a
missing ``input infection'' and reaching the terminal vertices
requires all $K+1$ input infections since $f(1/2) = 0$ (each chain
vertex has two inputs, and if either is uninfected, the chain of
infections halts). Similarly, if either player attempts to defect by
invading the seeded nodes of the other player, there will be
no payoff for either player in any of the replications. Thus the
most Blue can obtain by deviation is $N$ (moving its one infection
to the final chain vertex of a single replication), while the most
Red can obtain is $KN$ (moving all of their infections to the final
chain vertices of $K$ replications. The equilibrium requirements are
thus $M(N/2^K) > N$ for Blue, and $MN(1-1/2^k) > KN$ for Red. The
Blue requirement is the stronger one, and yields $M > 2^K$. The Budget Multiplier
for this configuration is the same as for the single replication
case, and thus if we let $K$ be as large as desired and choose $M >
2^K$, we can make the Budget Multiplier exceed any value.
\qed\ (Theorem~\ref{theoremPOB3})
\bigskip

It is worth noting that even if the Blue player has $L > 1$ seeded nodes, and we
repeat the construction above with chain length $K + L - 1$, but with Blue forced to
play at the beginning of the chain, followed by all the Red infections, the argument
and calculations above are unchanged: effectively, Blues $L$ seeded nodes are no
better than 1 infection, because they are simply causing a chain of $L-1$ Blue infections
before then facing the chain of $K$ Red inputs. In fact, even if we let $L \gg K$, Blue's
payoff will still be a factor of $1/2^K$ smaller than Red's. Thus in some sense the
theorem shows that if $f$ is sufficiently convex, not only is the Budget Multiplier unbounded, but
the much {\em smaller\/} initial budget may yield arbitrarily {\em higher\/} payoffs!

\section{Concluding Remarks}

We have developed a general framework for the study of competition between firms who use their resources to maximize adoption
of their product by consumers located in a social network. This framework yields a very rich class of competitive strategies,
which depend in subtle ways on the dynamics, the relative budgets of the players and the structure of the
social network. We identified properties of the dynamics of local adoption under which resource use by players is
efficient or unboundedly inefficient. Similarly, we identified adoption dynamics for which networks neutralize or dramatically accentuate
ex-ante resource difference across players.

There are a number of other questions which can be fruitfully investigated within our framework.
One obvious direction is to understand the structure of equilibria in greater detail, and in
particular how it is related to network structure. While our results on the PoA and Budget Multiplier demonstrate
that network structure can interact in dramatic ways with the switching and selection
functions at equilibrium, a more general and detailed understanding would be of interest.

Our model assumed that  players' budgets are exogenously given. In many contexts, the
budget may itself be a decision variable. It is important to understand if endogenous budgets would aggravate
or mitigate the problem of high PoA. Large network advantages from resources (reflected in high Budget Multiplier)
create an incentive to increase budgets, and may be self-neutralizing.
\iffalse
To see how endogenous budgets can have a large impact, consider the case where switching function is concave and
selection function is linear. Suppose a player can purchase one unit of resource at cost $c=1/2$. The final payoffs to a player
are then equal to the number of adoptions less the seed expenditures. When the network is a star with a single hub,
it is an equilibrium for both players to buy $n/2$ units  and locate at the hub, thus yielding a payoff $0$.
By contrast, the joint payoffs are maximized with one unit of resource and yield total joint payoffs
$n-1/2$. The PoA is unbounded.

Similarly, the order of moves can have a large impact in certain networks.
Suppose Red moves first and has budget 1, while Blue moves second and has budget 2.
The switching function and selection function are linear and a consumer is perpetually active until he
adopts. In the ring network, Red will earn 1, while Blue will earn $n-2$. The Budget Multiplier can be
made arbitrarily large by raising the value of $n$. By contrast, in the star network,
the Budget Multiplier is equal to 1, irrespective of the order of moves of the players.
\fi

Other interesting directions include algorithmic issues such as computing equilibria and best
responses in our framework, and how their difficulty depends on the switching and selection functions;
and the multi-stage version of our game, in which the two firms may gradually spend their seed budgets
in a way that depends on the evolving state of the network.

\section*{Acknowledgements}
We warmly acknowledge Hoda Heidari for many insightful comments and discussions, and her
extensions of our results mentioned in the paper.
We give special thanks to David Kempe for many valuable comments and suggestions.
We also thank Andrea Galeotti, Rachel Kranton, Tim Roughgarden, Muhamet Yildiz and seminar participants at
Essex, Cambridge, Microsoft Research (Cambridge), Queen Mary,
Warwick and York for useful comments.

% \newpage

\newpage
\appendix

One criticism of the model presented here is that while we assume rationality on the part of the
competing firms, consumers (represented by the vertices in the social network) behave in a purely
stochastic, non-rational fashion. In this appendix, we sketch natural cases in which these
stochastic decisions can actually be provided with rational microeconomic models.
We illustrate how the switching and selection functions $f$-$g$ may be founded upon
optimal decisions made by consumers located in social networks. Information sharing about products and
direct advantages accruing from adopting compatible products are two important ingredients
of local social influence.

\textbf{Example 1: Information Sharing:} Consumers are looking for a good whose utility depends on its quality;
the quality is known or easily verified upon inspection (such products are referred to as `search' goods),
but its availability may not be known. Examples of such products might be televisions and desktop computers.
Consumers search on their own and they also get information from their friends and neighbors.
Suppose for simplicity that the consumer talks to one friend before making his decision.
As he runs into friends at random, other things being equal,
the probability of adopting a product is equal to the
probability of meeting someone who has adopted it. This probability is in turn given by
the fraction of neighbors who have adopted the product.
This corresponds to the case where $f$ and $g$ are both linear.

\textbf{Example 2: Information Sharing and Payoff Externalities.}
Consumers are choosing between goods whose utility depends on how many other consumers have adopted the
same product. Prominent examples include social networking sites.
Suppose products offer stand alone advantage $v$, and a adoption related reward which
is equal to the number of Reds or the number of Blues. Consumer picks neighbors at
random. If neighbor is Red or Blue, then consumer becomes aware of
product market. There is a small cost (relative to $v$) at which he
can ask all his neighbors about their status.
He then compares the adoption rates of the two products and given the
benefits to being in a larger (local) network, the consumer selects the more popular product.
This situation gives rise to an $f$ which is increasing and concave in the fraction of adopters,
while $g$ is polarizing (close to a winner-take-it all).

We leave to future work the formulation of further and richer microeconomic consumer models,
including a fully game-theoretic formulation over both firms and consumers.
\end{document}